\documentclass[preprint,apj]{emulateapj}

\newcommand{\kms}{{km s$^{-1}$}}

\newcommand{\Hb}{H$\beta$~}
\newcommand{\Hbb}{H$\beta$}
\newcommand{\Ha}{H$\alpha$~}
\newcommand{\Hab}{{H$\alpha$}}
\newcommand{{\Stwo}}{{[\ion{S}{2}]}}
\newcommand{\Oone}{{[\ion{O}{1}]}}
\newcommand{\Otwo}{{[\ion{O}{2}]}}
\newcommand{\Othree}{{[\ion{O}{3}]}}
\newcommand{\Ntwo}{{[\ion{N}{2}]}}
\newcommand{\Htwo}{{\ion{H}{2}}~}
\newcommand{\HtwoR}{{\ion{H}{2}~ region}}

\newcommand{\chb}{{$c$(H$\beta$)}}

\usepackage{epstopdf}
\usepackage{multirow}
\usepackage{color}

\shorttitle{SNRs in M81 and M82}
\shortauthors{Lee et al.}

\begin{document}

\title{Optical Spectroscopy of Supernova Remnants in M81 and M82}

\author{Myung Gyoon Lee$^{1}$, Jubee Sohn$^{1}$, Jong Hwan Lee$^{1,2}$, Sungsoon Lim$^{1,3,4}$, In Sung Jang$^{1}$, Youkyung Ko$^{1}$, Bon-Chul Koo$^{1}$, Narae Hwang$^{5}$, Sang Chul Kim$^{5,6}$, \& Byeong-Gon Park$^{5,6}$}
\affil{$^{1}$Astronomy Program, Department of Physics and Astronomy, Seoul National University, Gwanak-gu, Seoul 151-742, Korea}
\affil{$^{2}$Aerospace Research Center, Korea Air Force Academy, Cheongju 363-849, Korea} 
\affil{$^{3}$Department of Astronomy, Peking University, Beijing 100871, China}
\affil{$^{4}$Kavli Institute for Astronomy and Astrophysics, Peking University, Beijing 100871, China}
\affil{$^{5}$ Korea Astronomy and Space Science Institute, Daejeon, 305-348, Korea}
\affil{$^{6}$ Korea University of Science and Technology (UST), Daejeon, 305-350, Korea}
\email{mglee@astro.snu.ac.kr}

Recieved 2014 December 24; accepted 2015 February26


\begin{abstract}
We present spectroscopy of 28 SNR candidates as well as one \Htwo region in M81, and two SNR candidates in M82.
Twenty six out of the M81 candidates turn out to be genuine SNRs, and two in M82 may be shocked condensations in the galactic outflow or SNRs. 
The distribution of \Ntwo/\Ha ratios of M81 SNRs is bimodal. 
M81 SNRs are divided into two groups 
in the spectral line ratio diagrams: an \Othree-strong group 
and an \Othree-weak group. 
The latter have larger sizes, and may have faster shock velocity. 
\Ntwo/\Ha ratios of the SNRs show a strong correlation with  \Stwo/\Ha ratios.
They show a clear radial gradient in \Ntwo/\Ha and \Stwo/\Ha ratios:
dLog ({\Ntwo/\Hab})/dLog R $ = -0.018\pm0.008$ 
dex kpc$^{-1}$  and dLog ({\Stwo/\Hab})/dLog R $=-0.016\pm0.008$ dex kpc$^{-1}$ 
where $R$ is a deprojected galactocentric distance.  
We estimate the nitrogen and oxygen abundance of the SNRs from the comparison with shock-ionization models.
We obtain a value for the nitrogen radial gradient, dLog({N/H})/dLogR $ = -0.023\pm0.009$ dex kpc$^{-1}$, and little evidence for the gradient in oxygen. 
This nitrogen abundance 
shows a few times flatter gradient than those of the planetary nebulae and \Htwo regions.
We find that five SNRs
are matched with X-ray sources. Their X-ray hardness colors are consistent with thermal SNRs.
\end{abstract}
\keywords{galaxies: spiral  --- galaxies: individual (M81, M82)  --- galaxies: ISM--- galaxies: abundances }

\section{Introduction}
Supernova remnants (SNRs) in nearby galaxies are an ideal target to study various statistical aspects of SNRs, 
because we can find them in the entire region of a galaxy (within the survey limits), 
 and because they are in the same distance from us.  
SNR candidates in nearby galaxies are often found using narrow band imaging with \Ha and \Stwo~ filters, X-ray imaging, and radio observation
 (see \citealp{lon85,mat97,pan07,leo10,lee14a,lee14b} and references therein). 
Optical spectroscopy of the SNR candidates is a powerful tool
 to confirm whether they are genuine SNRs or not,
 to derive their physical parameters (electron densities, temperatures, shock velocities)
 and chemical abundances, and
 to investigate the radial gradient of the abundances \citep{dop77,ray79,shu79,dop84}.
Optical spectroscopic studies of SNRs in several nearby galaxies have been done in the past.
The Local Group galaxies have been primary targets of optical spectroscopic studies:
 LMC, SMC \citep{rus90,pay08},
 M31 \citep{bla81,bla82,gal99}, and
 M33 \citep{bla85,smi93,gor98}. 
Recently SNRs in several low luminosity disk galaxies have been studied spectroscopically \citep{leo13}.

\citet{gor98} performed an extensive optical study for a large sample of SNRs in M33 including those in the previous studies \citep{bla85,smi93}.
They presented several interesting results on the correlation between line ratios:
 no relation between \Ntwo/\Ha and \Stwo $\lambda$6717/\Stwo $\lambda$6731,
 a weak relation between the SNR diameter and \Ntwo/\Ha as well as \Stwo/\Hab, and
 a strong correlation between \Ntwo/\Ha and \Stwo/\Hab, and between \Oone/\Ha and \Stwo/\Hab.
They found also that 
 the line ratios of \Ntwo/\Ha of M33 SNRs decrease as the galactocentric distance (GCD, $R$) increases at $0<R<6$ kpc,
 and they show a significant dispersion at a given galactocentric distance. 
They suggested that
 the significant dispersion is mainly due to a larger abundance dispersion in the inner region of the galaxy. 
However, their spectra covered only the red region, 6200 to 7500 \AA~ so that they could not have any information of the blue emission lines.
exi
Optical spectroscopic studies of the SNRs in M31 were given by \citet{bla81,bla82,gal99}.
From the analysis of the combined sample including those in the previous studies \citep{bla81,bla82}, \citet{gal99} 
found a clear radial gradient of \Ntwo/\Ha in the range of $3 <R< 20$ kpc ($-0.04\pm0.01$ dex kpc$^{-1}$ in their Fig. 6d),
and little evidence for gradients in \Stwo/\Hab, \Othree/\Hbb, \Othree/\Otwo, and electron temperature ($T_e$).
They found also a significant correlation between \Othree/\Hb and \Otwo/\Hbb, noting that it may be mainly due to oxygen abundance variation.

M81 (NGC 3031) is another nearby bright spiral (SA(s)ab) galaxy, located beyond the Local Group.
It is a main member of the M81 Group, interacting with the well-known starburst galaxy M82 and a dwarf galaxy NGC 3077 with morphology type of I0. 
\citet{mat97} provided a list of 41 SNR candidates selected using the criterion
 of \Stwo/\Ha $\geq 0.45$ from the \Stwo~ and \Ha images of M81. 
However, they obtained optical spectra of only four in their sample (MF No. 2, 17, 18, and 25)
 covering 4800--7200 \AA, and confirmed that they are indeed SNRs.

In this study we present a spectroscopic study of  a large number of SNR candidates in M81, 
 using a wide spectral range to cover from \Otwo 3727 to \Stwo $\lambda$6731 lines. 
This paper is composed as follows. 
Section 2 describes how to select and observe the targets and how to reduce the spectroscopic data.
In \S 3 we classify SNRs, investigate the property of emission line ratios of the SNRs, and compare them with shock-ionization models. Then we investigate the relations between the line ratios and sizes of the SNRs, and any radial variation of the line ratios and abundances.
In \S4 we discuss the radial gradient of chemical abundances derived from the SNRs in comparison with those based on \Htwo regions and planetary nebulae (PNe) in M81. We compare the radial gradients of M81 SNRs with those of the SNRs in other nearby galaxies.  Also we match optical SNRs with X-ray source catalogs of M81.  
Final section summarizes the main results with conclusions.

We adopted a distance to M81, $3.63\pm0.14$ Mpc \citep{dur10},
 inclination angle, $i=58$ deg, and position angle, $PA=157$ deg to calculate the deprojected galactocentric distances ($R$) 
 of the targets \citep{dev91,pue14}.
The standard radius ($R_{25}$) of M81 is 18.34 kpc and its heliocentric radial velocity is $-34\pm4$ \kms~ \citep{dev91}. 
One arcsec corresponds to 17.6 pc at the distance of M81.
The distances to M81 and M82 are similar \citep{lim13}.
 
\section{Observations and Data Reduction}

\subsection{Target Selection}

Primary targets (20) were selected from the list of the SNR candidates in M81 given by \citet{mat97}. 
We noted their IDs starting with MF.
Then we added eight secondary targets we selected as SNR candidates from the \Ha and \Stwo $\lambda\lambda$6716,31 images
 obtained with INT 2.5m in the Isaac Newton Group Archive.
Their IDs run from L1 to L8.
We also selected one \Htwo region in the southeast region of M81 for reference.
 
\begin{figure}
\centering
\includegraphics[scale=0.35]{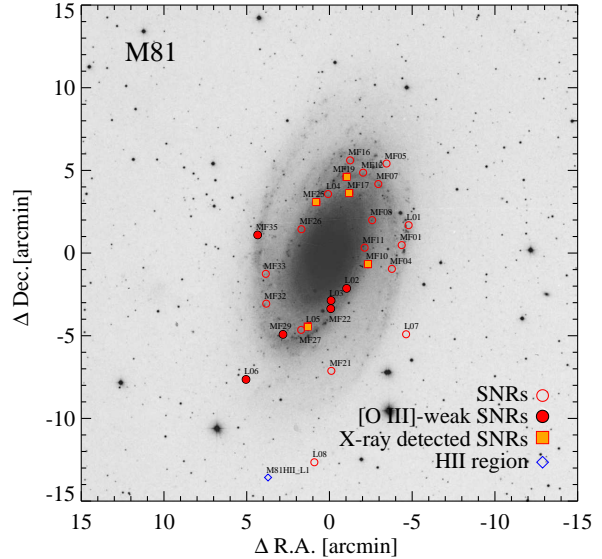}
\caption{
The spatial distribution of the SNR candidates (circles) and one \Htwo region (diamond) in M81 in this study, 
 marked in the grayscale map of the digital sky survey. 
North is up and east to left. Filled circles represent \Othree-weak SNRs (with \Othree/\Hb$<1.0$),
 and filled squares SNRs matched with X-ray sources \citep{sel11}. }
\label{m81finder}
\end{figure}

\begin{figure}
\centering
\includegraphics[scale=0.35]{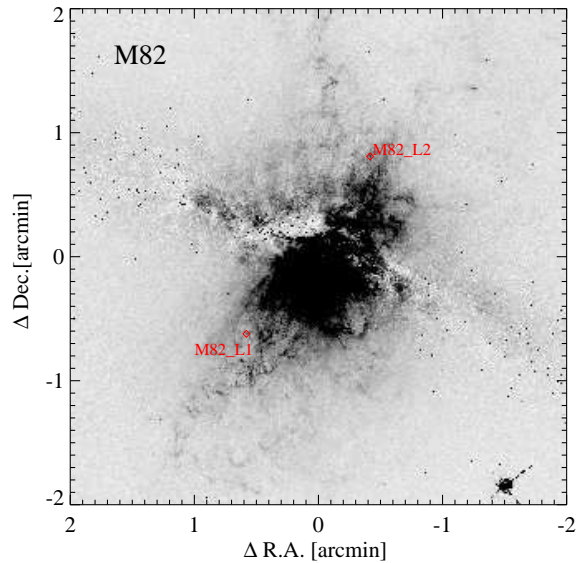}
\caption{Location of two SNR candidates in M82 (M82L1 and M82L2) in this study, overlayed in the 
gray scale map of continuum-subtracted ACS \Ha image in the HST archive (PI : Mountain, PID : 10776).
North is up and east to left. Note that they are located in the outflow direction.}
\label{m82finder}
\end{figure}

In addition, we checked the \Ha and \Stwo~ images of M82 in the Hubble Space Telescope (HST) archive
 (PI: Mountain and PID: 10776 for \Ha images, and PI: O'Connell and PID: 11360 for \Stwo~ images). 
\citet{deg00} presented a list of 10 SNR candidates in the northeast starburst region (M82B) of M82  that are compact \Ha-bright sources,
 selected from the \Ha images obtained with HST/WFPC2. 
There are no spectroscopic studies of these candidates so that their nature remain to be known. 
Similarly we selected two compact \Ha-bright objects located in the outflow region of M82 as interesting targets. 
{\bf Figures \ref{m81finder} and \ref{m82finder}} display the location of the targets in M81 and M82, respectively. 
Note that the M81 targets are mostly located along the spiral arms.

\subsection{Observation}

To obtain the spectra of the targets we used the Hectospec fiber-fed spectrograph (Hectospec) 
 on the 6.5m Multi Mirror Telescope (MMT)\citep{fab05}.
We selected a 270 mm$^{-1}$ grating at a dispersion of 1.2 \AA~ pixel$^{-1}$,
 covering 3650 \AA~ to 9200 \AA~ at a spectral resolution of about 5 \AA. 
However, the sensitivity at the blue and red limits are low so that we used only
 the spectral range of  3700 \AA~ to 7000 \AA~ for analysis.
The spectrograph feeds 300 fibers over one degree diameter field of view
 and the diameter of each fiber is $\sim1\farcs5$, corresponding to  26.4 pc at the distance to M81.

The observations of the emission line objects in M81 and M82 were carried out as a part of our larger M81 program in queue mode 
during the months of February to May 2014.
We took 3 to 5 exposures of 1440 s for one configuration. 
We observed four configurations that are overlapped partially with each other.
Total exposure times for one configuration are 4320 s to 7200 s.
Seeing values during the observations ranged from 1\farcs0 to 1\farcs2. 
 
\subsection{Data Reduction}

We reduced the data using HSRED version 2 written by R.Cool\footnote{http://mmto.org/rcool/hsred/} \citep{koc12},
 including debiasing, flat-fielding, aperture extraction of spectra, wavelength calibration, and flux calibration. 
Sky subtraction for spectra extraction was done using the average of the spectra for the nearest blank-sky fibers. 

We measured the emission-line fluxes using Gaussian fitting after continuum subtraction in IDL.
The errors for the fluxes were estimated including the systematic errors of the flux calibration, background estimation, and sky subtraction.
For extinction correction we derived \chb, the logarithmic extinction at \Hb, 
 adopting the intrinsic value of 
 \Ha/\Hb = 3.0 for the SNRs based on the shock models \citep{ray79,shu79} and 
 2.86 for HII regions based on Case B recombination and a gas with $T_e \approx 10^4$ K \citep{ost06}. 
We adopted the extinction curve by \citet{car89}.

We measured the radial velocities of the targets using HSRED version 2,
 after checking visually whether each measurement is reasonable or not.
We adopted the range of 3900\AA~ to 6400 \AA~ for velocity measurements.
The mean value of the measurement errors of the radial velocities is 3 \kms.
Table \ref{line} lists a list of spectroscopic parameters of the targets:
 ID, $R$ [kpc], radial velocity with its error, \chb, \Hb strength, and the extinction corrected line fluxes and errors, which are normalized by \Hb$=100$. 
Hereafter we use only the extinction-corrected values for the line fluxes and line ratios for the following analysis.

One of the secondary targets (L7) showed too poor a spectrum to use. 
Another of the secondary targets (L8) turned out to be a background galaxy at $z=0.33$. 
Thus we used the spectra of 29 objects for analysis.
Two of our sample are matched with the SNRs for which spectroscopic data are given in \citet{mat97}: MF17 and MF25.
The line ratios of these two SNRs in this study and \citet{mat97} agree well within the differences of about 20\%.

\section{Results} 

\subsection{SNR Classification}

\begin{figure}
\centering
\includegraphics[scale=0.5]{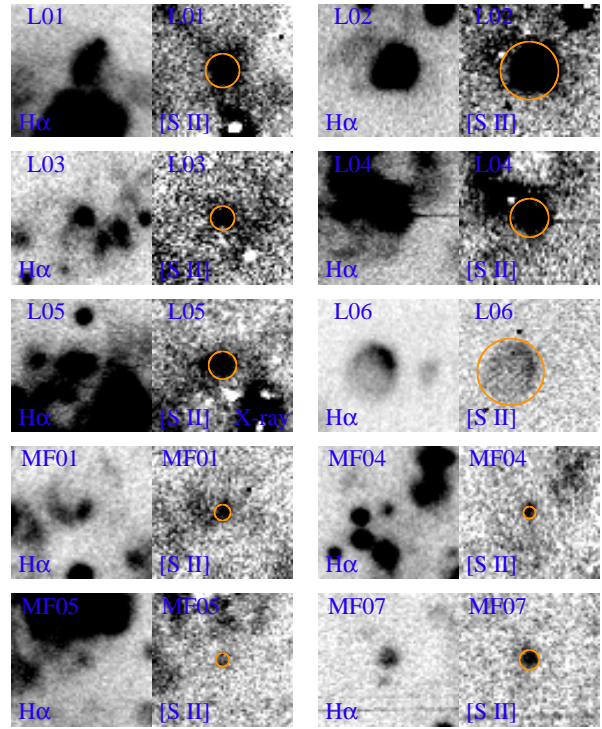}
\caption{Stamp maps  ($20\arcsec \times 20\arcsec$) of the SNRs and one \Htwo region in M81 in this study,
 made from the continuum-subtracted \Ha and \Stwo images. 
North is up and east to the left. 
Circles represent the SNR size estimated in this study. 
Note that the fiber with diameter $1\farcs5$ was located in the center of each map.
Five SNRs are matched with the X-ray source catalog \citep{sel11}:
MF10=Sell259, MF17=Sell193, MF19=Sell195=Liu1578, MF25=Sell172, and L5=Sell50=Liu1653.}
\label{snrmap}
\end{figure}

\begin{figure}
\centering
\includegraphics[scale=0.5]{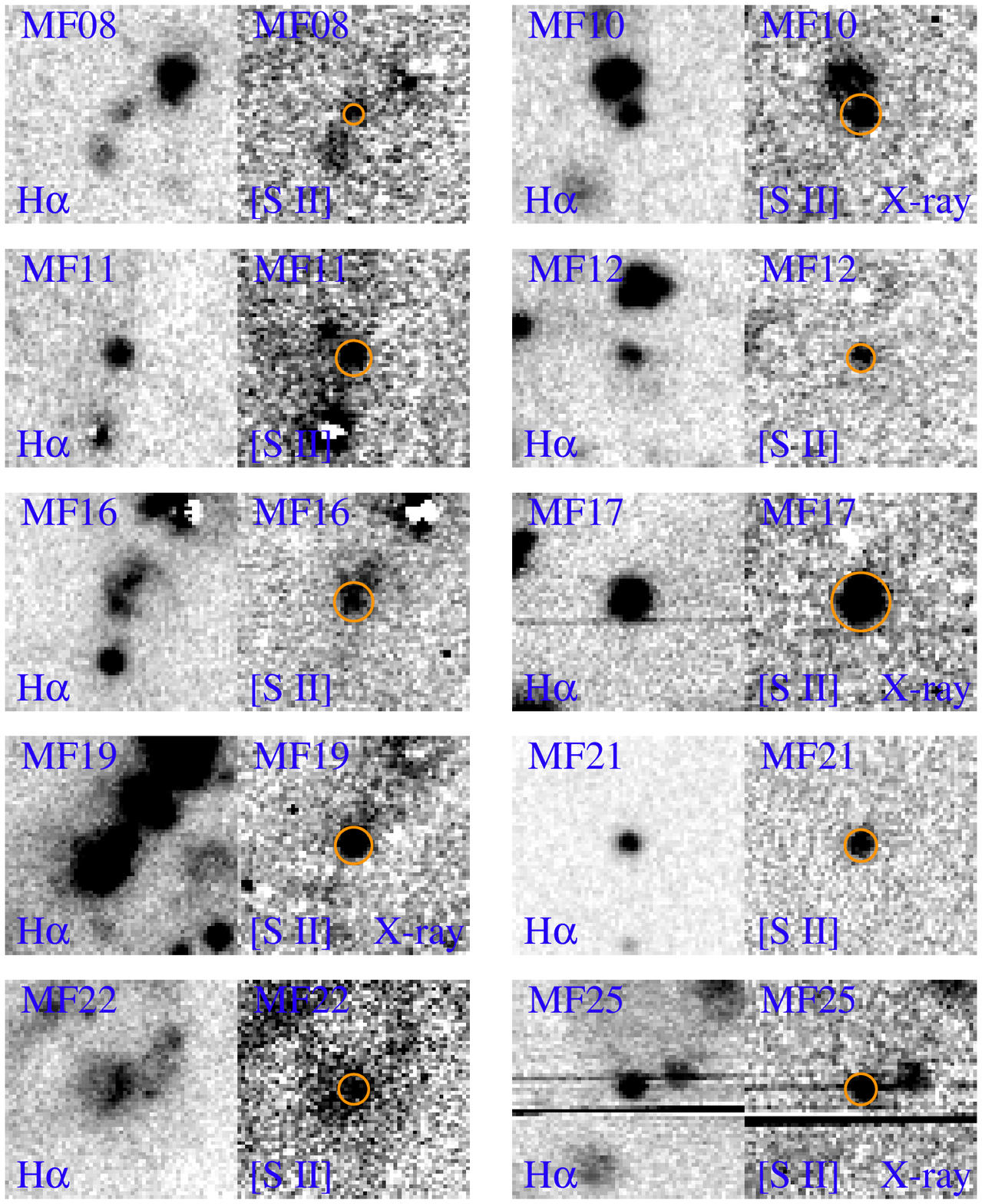} 
\figurenum{3}
\caption{Continued.}
\end{figure}

\begin{figure}
\centering
\includegraphics[scale=0.5]{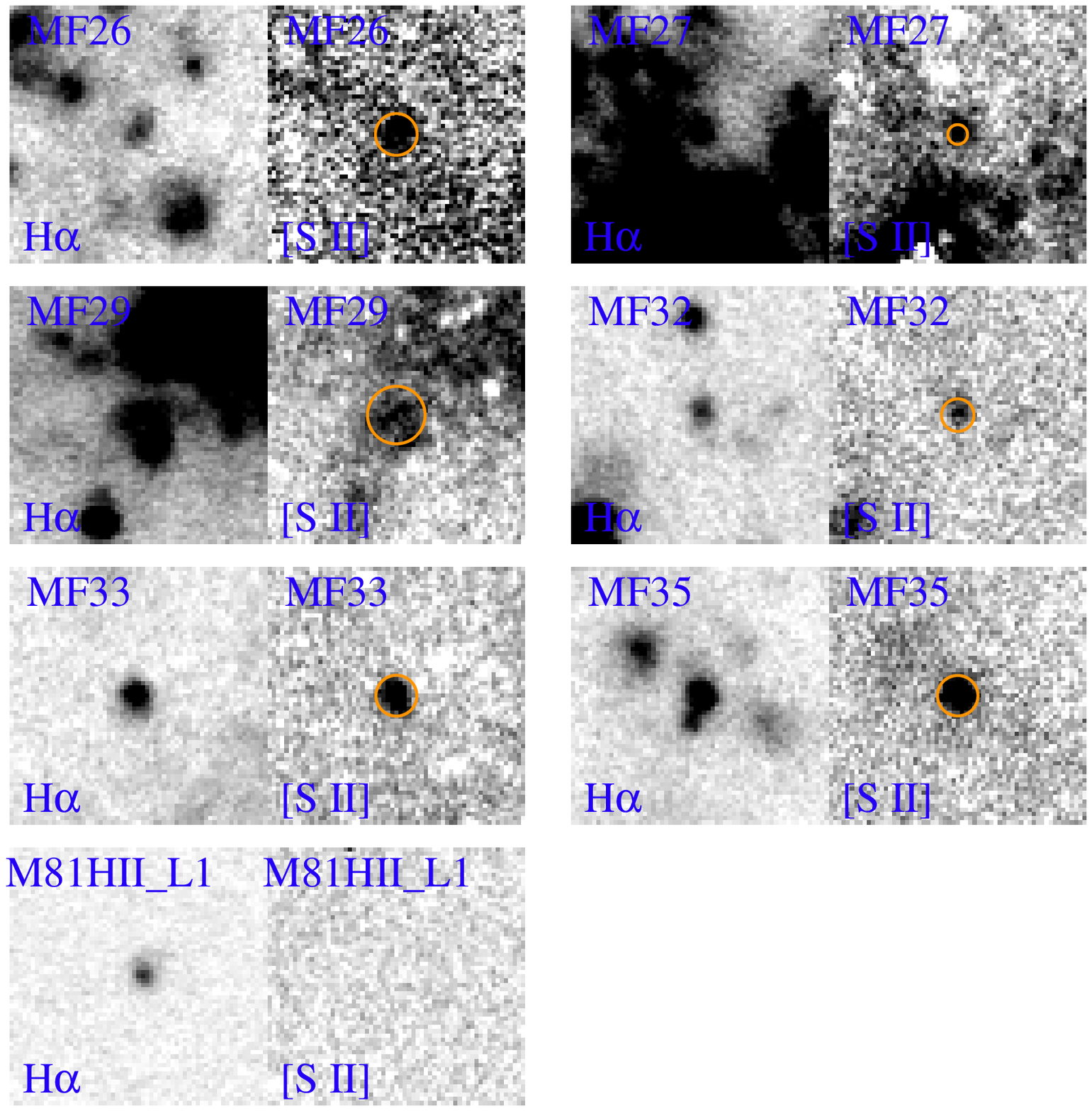} 
\figurenum{3}
\caption{Continued.}
\end{figure}

\begin{figure}
\centering
\includegraphics[scale=0.5]{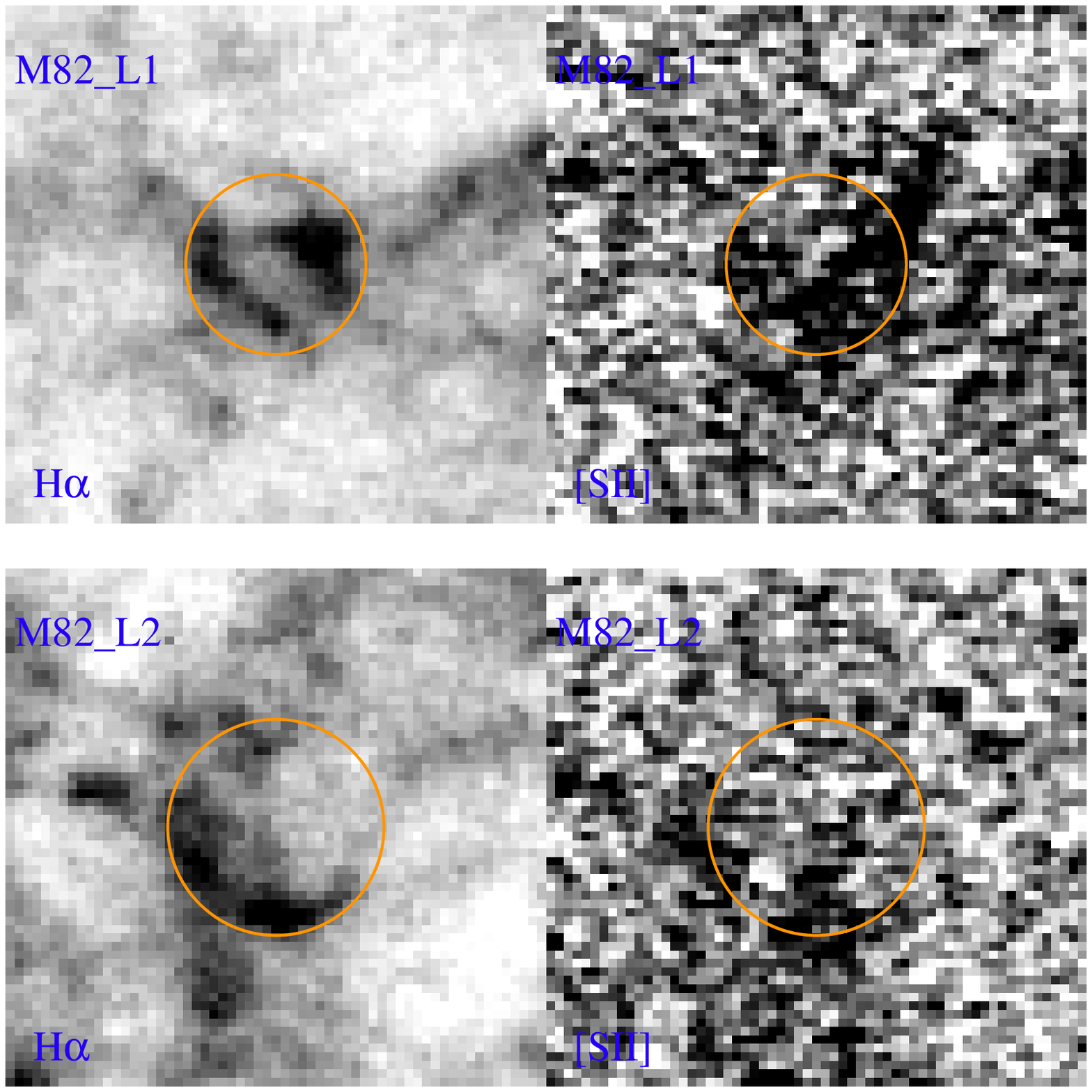}
\caption{Same as Figure \ref{snrmap} but for  M82 SNRs.
Each map covers $6\arcsec \times 6\arcsec$, 
 made from the continuum-subtracted ACS \Ha and \Stwo~ images.
Circles represent the size of the SNRs.
Note that they show shell structures typical for SNRs.} 
\label{m82map}
\end{figure}

{\bf Figure \ref{snrmap}} displays stamp maps of the SNR candidates and one \Htwo region
 in M81, based on the continuum-subtracted \Ha and \Stwo~ images we prepared using the INT images. 
We estimated visually the diameters of the SNR candidates in the continuum-subtracted \Stwo~ images,
 as marked by circles in Figure \ref{snrmap},
 and listed them in Table \ref{ratio}.
 Note that the contrast between the \Ha and \Stwo~ images is much higher for the \Htwo region than for the SNR candidates.  
In {\bf Figure \ref{m82map}} we showed stamp maps of two M82 SNR candidates,
 based on the continuum-subtracted \Ha and \Stwo~ images we prepared using the ACS \Ha and \Stwo~  images in the HST archive.
 Note that the high resolution images of these two objects show shell structures typical for SNRs or shocked regions.

\begin{figure*}
\centering
\includegraphics[scale=0.45]{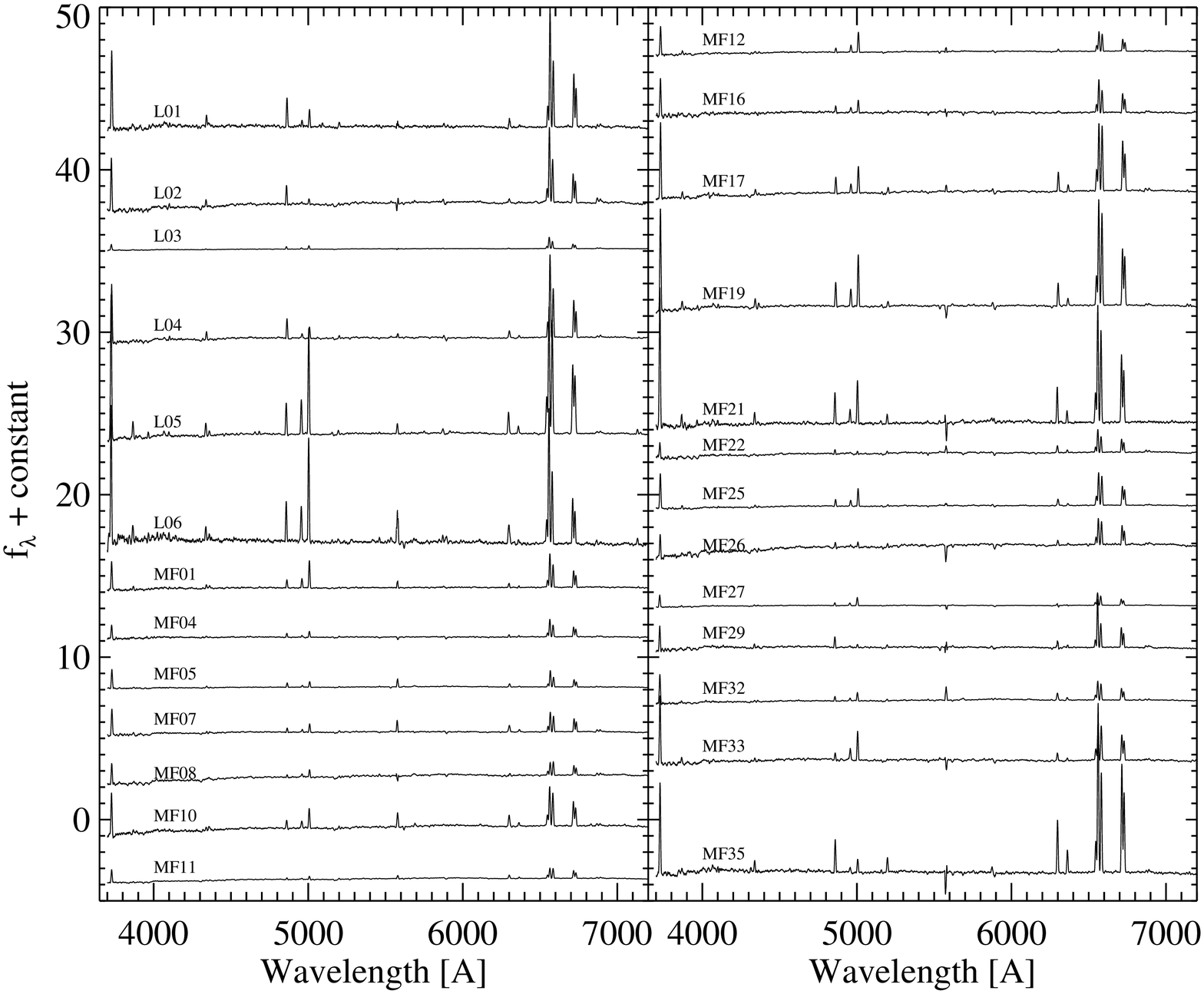}
\caption{Spectra of M81 SNRs in this study. 
Note several prominent lines typical for SNRs: \Otwo $\lambda\lambda$3727,9, \Hbb, \Othree $\lambda\lambda$4959,5007,
 \Oone $\lambda$6300, \Ntwo $\lambda$6548,83, \Hab, and \Stwo $\lambda\lambda$6717,31.}
\label{snrspec}
\end{figure*}

\begin{figure}
\centering
\includegraphics[scale=0.4]{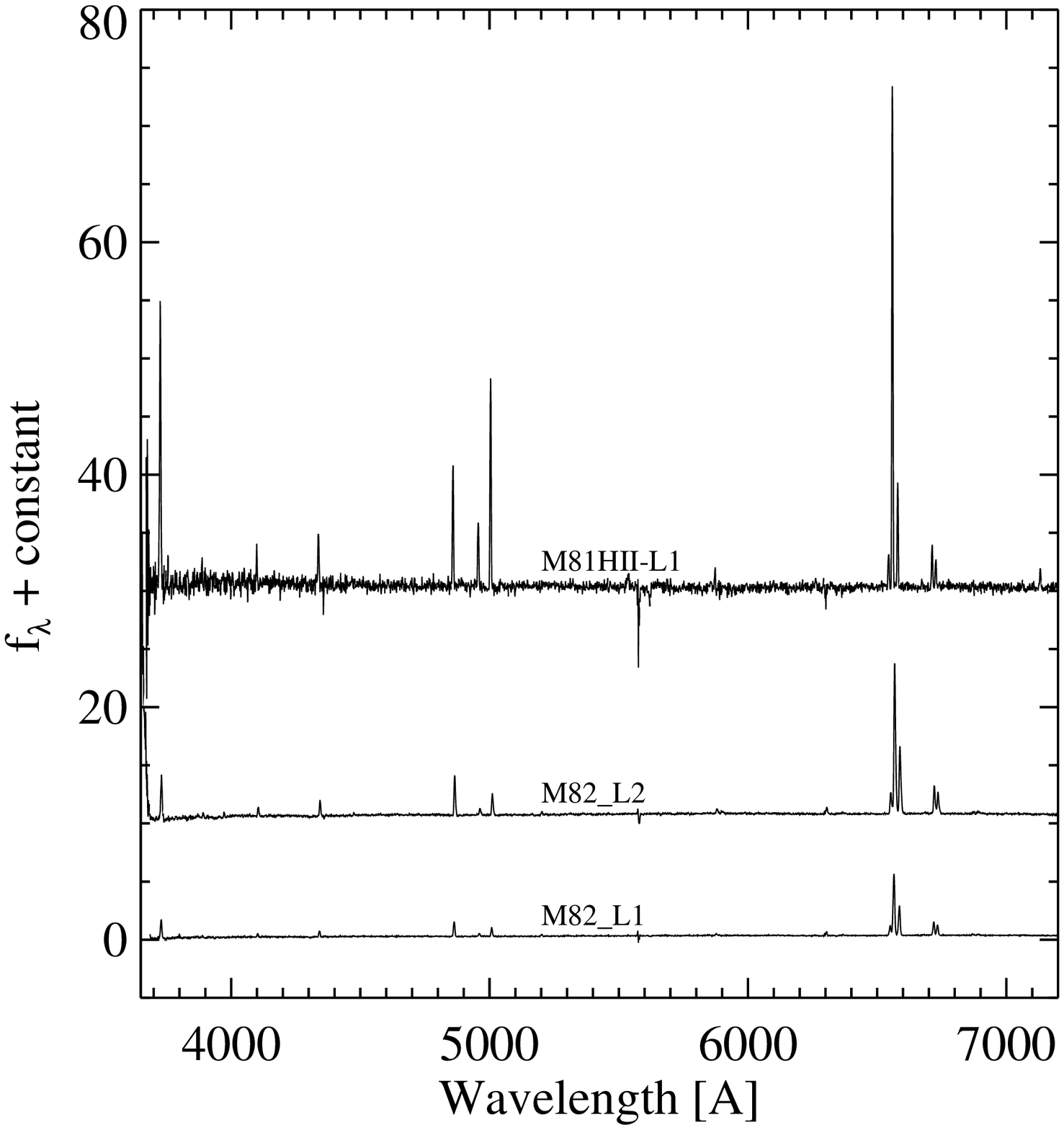} 
\caption{Spectra of one \Htwo region in M81 and two SNR candidates in M82. }
\label{h2spec}
\end{figure}
 
{\bf Figures \ref{snrspec} and \ref{h2spec}} show spectra of 26 M81 SNR candidates, 
two M82 SNR candidates and one M81 \Htwo region, respectively. 
Note several prominent lines typical for SNRs:
 \Otwo $\lambda\lambda$3727,9, \Hb, \Othree $\lambda\lambda$4959,5007, \Oone $\lambda$6300, \Ntwo $\lambda\lambda$6548,83, \Ha, and \Stwo $\lambda\lambda$6717,31.
\Othree $\lambda$4363 lines are too weak to measure for all objects.
Hereafter we use  for common doublet lines,
\Otwo~ for \Otwo $\lambda\lambda$3726,9, 
\Othree~ for \Othree $\lambda\lambda$4959,5007, 
\Ntwo~ for \Ntwo $\lambda\lambda$6548,83, and
\Stwo~ for \Stwo $\lambda\lambda$6716,31.

\begin{figure}
\centering
\includegraphics[scale=0.5]{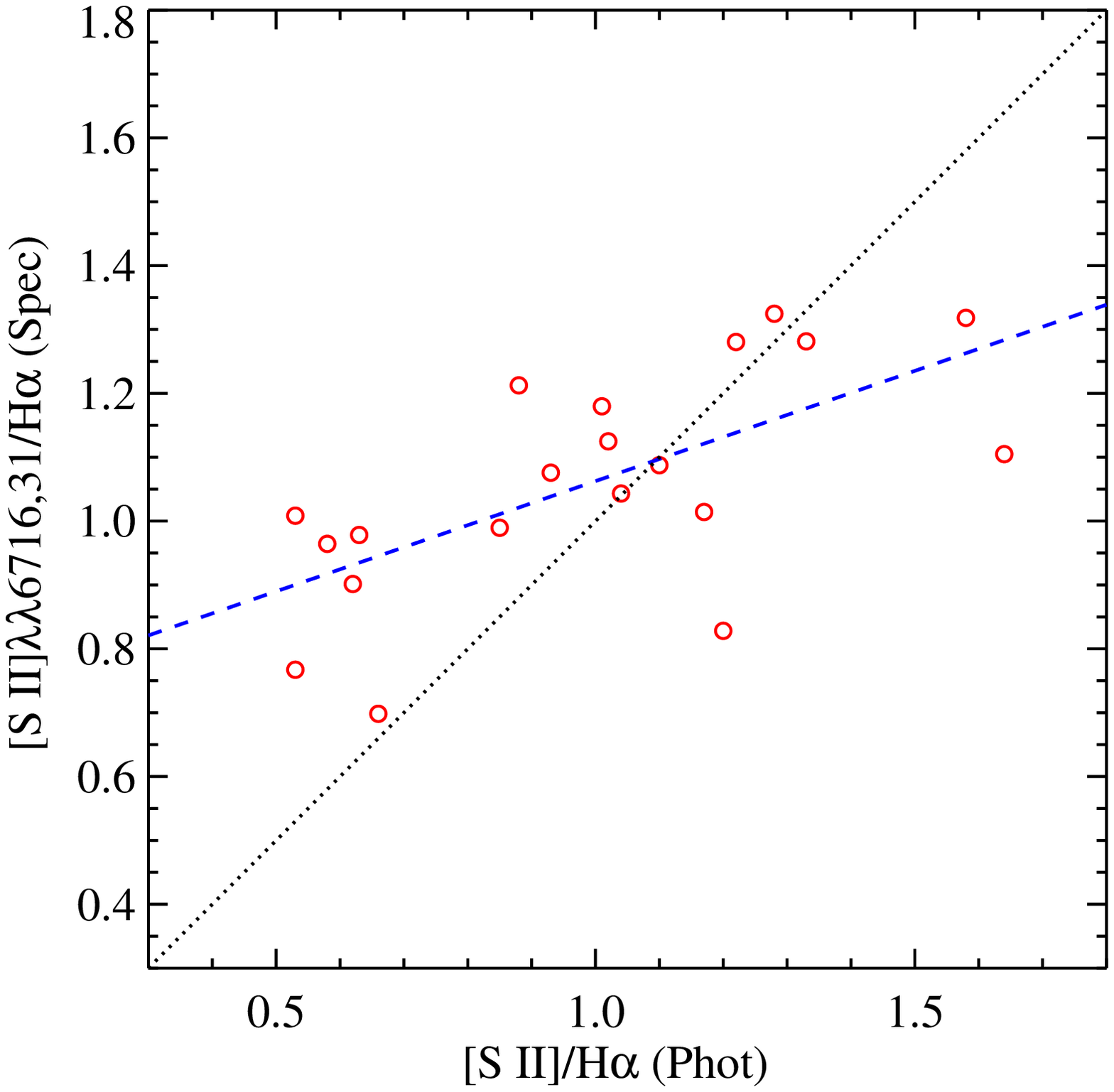} 
\caption{Comparison of \Stwo/\Ha ratios
of M81 SNRs derived from photometric images
and spectroscopic data in this study.
The dashed-line and dotted-line represent one-to-one relation and the best linear-fitting relation, respectively. }
\label{s2comp}
\end{figure}

The \Stwo/\Ha ratio has been considered to be an excellent criterion to distinguish SNRs from \Htwo regions \citep{lon85}. 
We derived the values of the \Stwo/\Ha ratio of M81 SNR candidates using their fluxes measured in the continuum-subtracted \Ha and \Stwo~ images.
In {\bf Figure \ref{s2comp}}, we compared thus photometrically derived values with those we measured from the spectra.
It shows a reasonable correlation between the two values, but with a large scatter. 
Linear-least-squares fitting yields
 \Stwo/\Ha (spec) = ($0.894\pm0.059$) \Stwo/\Ha (phot) ($0.121\pm0.061$) with $rms=0.084$,
 showing that its slope is slightly smaller than one.  
This confirms that the photometric method based on \Stwo~ and \Ha images to search for SNR candidates in nearby galaxies is an efficient tool.

\begin{figure}
\centering
\includegraphics[scale=0.425]{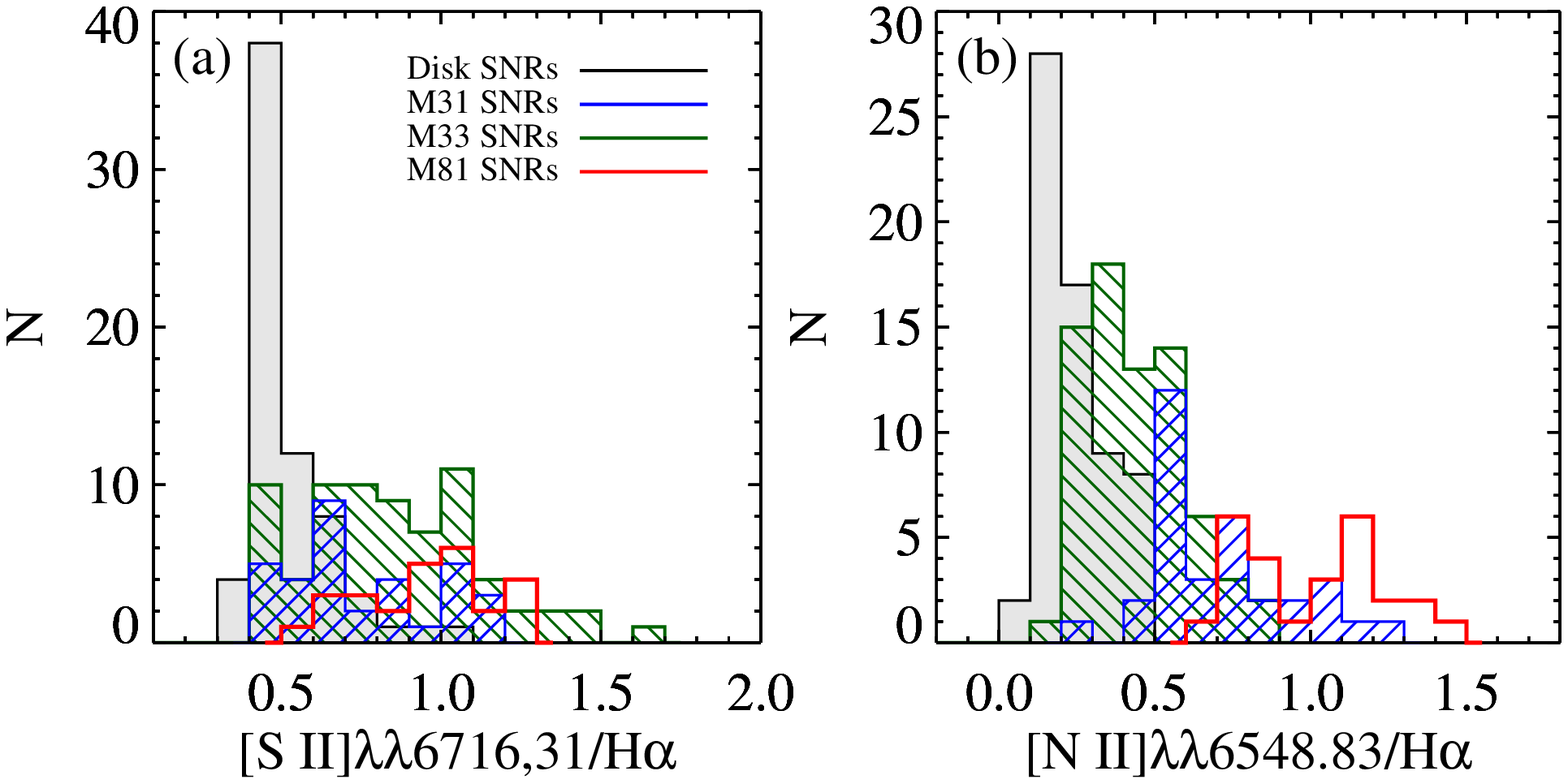}
\caption{The \Stwo/\Ha (a) and \Ntwo/\Ha (b)
 ratio distributions of M81 SNRs (this study)
 in comparison with M33 SNRs \citep{gor98}, 
 M31 SNRs \citep{gal99} and SNRs in six disk galaxies \citep{leo13}.
Note that the \Ntwo/\Ha ratio distribution of M81 SNRs is bimodal.}
\label{s2ha}
\end{figure}

We plotted the distributions of the \Stwo/\Ha and \Ntwo/\Ha ratios of M81 SNR candidates derived from spectra in 
{\bf Figure \ref{s2ha}}. 
We also plotted the data for M31 SNRs \citep{gal99}, M33 SNRs \citep{gor98}, and SNRs in six low luminosity disk galaxies \citep{leo13} for comparison. 
These figures show two notable features.
First, the \Stwo/Ha ratios of all 26 SNR candidates in M81 are higher than 0.5, showing that they are indeed SNRs.
Second, the \Ntwo/\Ha distribution of M81 SNRs is bimodal with peaks at 0.75 and 1.15. M31 SNRs also show a similar bimodal distribution.
However, the number ratio of the higher and lower components
 of M81 SNRs is larger than that for M31 SNRs.
It is noted that the presence of the higher component is mainly due to the SNRs located in the inner region at $R<5$ kpc, as shown later.

\begin{figure}
\centering
\includegraphics[scale=0.425]{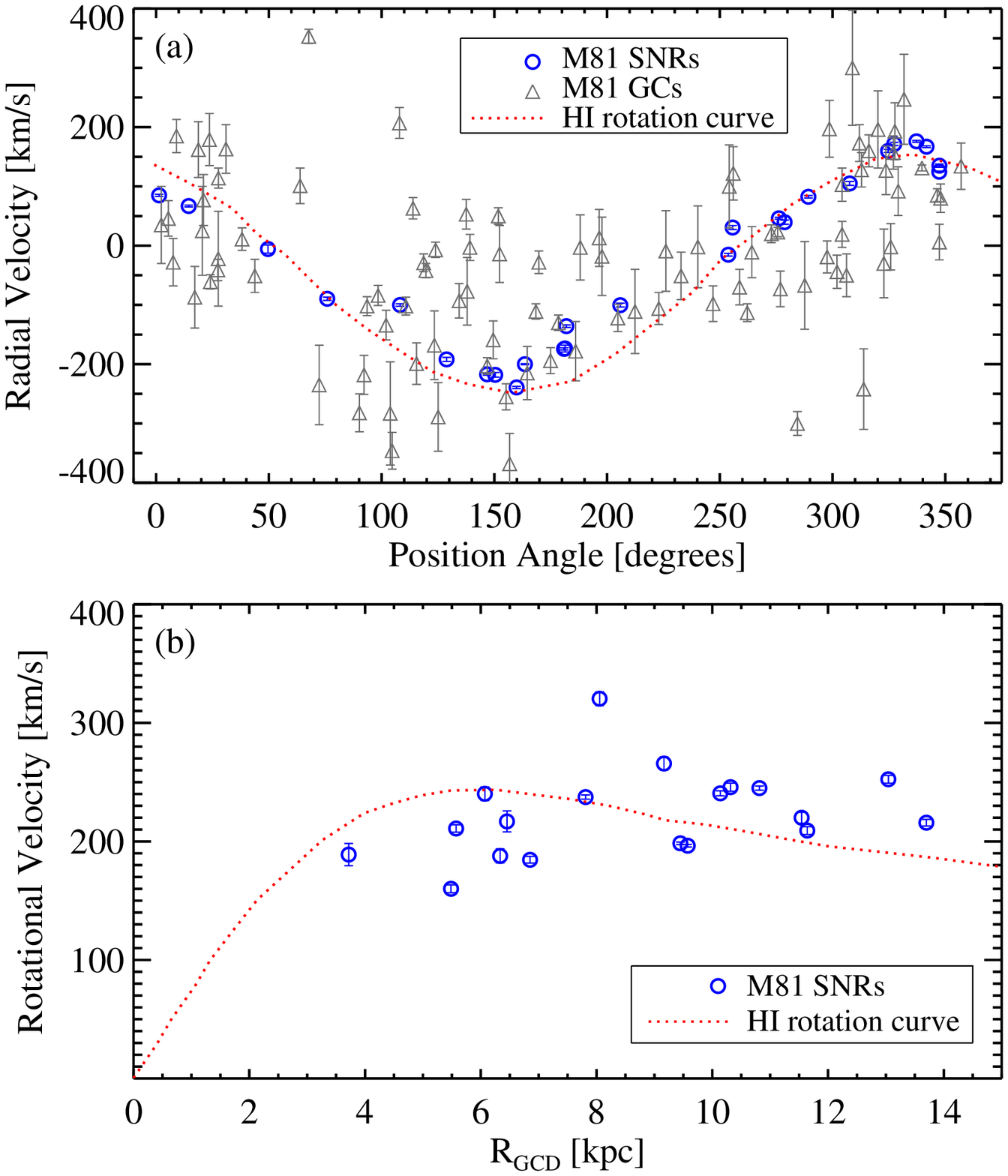}
\caption{Radial velocity with respect to the galaxy center of M81 SNRs (circles) vs. Position Angle (a),
 and corresponding rotational velocity versus galactocentric distance of the SNRs with position angles larger than 20 deg (b).
Triangles and crosses represent the data for HI gas \citep{rot75} and globular clusters in M81 \citep{nan10}.
Note that M81 SNRs follow well the rotation of the M81 disk.\\}
\label{rot}
\end{figure}

In Figure \ref{rot}(a), we displayed the radial velocity with respect to the galaxy center of M81 SNRs as a function of position angle, in comparison with those for HI gas \citep{rot75} and globular clusters \citep{nan10}. 
We derived the rotation velocity of the SNRs,  using
$v_c = (v_r - v_0)/\sin i \cos (\phi - \phi_0 )$ where $\phi$ and $\phi_0$ are position angles of the SNRs and the minor axis of M81, respectively, and $i$ is an inclination angle. 
We adopted the systemic velocity of M81, $v_0 = -34$ \kms \citep{dev91}.
We plotted the results for the SNRs with $(\phi - \phi_0 )>20$ deg as well as those of HI gas \citep{rot75} in Figure \ref{rot}(b). It is found that the rotational velocity data of M81 SNRs are in excellent agreement with those of HI gas, and are consistent with those of the globular clusters. This shows that all M81 SNRs in this study are located in the disk, following the disk rotation, so that they belong to the disk population of M81.

\subsection{Emission Line Diagnostics}

General characteristics of emission lines typical for shocked regions such as SNRs 
are described in relation with shocked-ionization models \citep{dop77,dop84, ray79,shu79, bla82,smi93,mat97,gor98,gal99,dop95,dop96,all08}.
We give only a brief summary of them as follows.
\Othree $\lambda$5007 lines are emitted in the region close to the shock front 
 so that their brightness is affected by the postshock electron temperature and the oxygen abundance.
The line ratio of \Othree $\lambda$5007/\Otwo $\lambda$3727 depends mainly on the postshock condition rather than on metallicity \citep{dop77}.
The line ratio of \Othree $\lambda\lambda$4959,5007/\Othree $\lambda$4363 is an indicator of electron temperature. 
\Othree $\lambda$4363 lines in SNRs are weak in general, although they are stronger than in \Htwo regions.
\Ntwo~ lines are emitted in the large recombination region behind the shock front,
 and are known to be an outstanding tracer of nitrogen abundance, little affected by shock temperature or electron density. 
Thus the line ratio of \Ntwo/\Ha is a good indicator of relative abundances of nitrogen and hydrogen \citep{dop84,smi93,gor98,gal99}. 
The ratio of the doublet lines, \Stwo $\lambda$6717/\Stwo $\lambda$6731, is a well-known indicator of electron density \citep{bla85}.
The ratio of \Stwo $\lambda$6731/\Ha is mainly sensitive to abundance, being little affected by shock conditions \citep{dop84,bla85,smi93}.
\Oone $\lambda$6300 lines become stronger as the shock velocity increases \citep{rus90}.
The ratios of these lines with respect to \Hb vary significantly depending on shock velocity in the low shock velocity range
 (if the shock velocity is smaller than $\sim 100$ \kms). 
However, some of them (\Otwo $\lambda\lambda$3727,9, \Othree $\lambda$5007, \Ntwo $\lambda$6584, and \Stwo $\lambda$6731) change little
 when the shock velocity is larger than $\sim 100$ \kms (see Figures 5 and 6 in \citet{dop84}) so that they can be used as an abundance indicator.

\begin{figure}
\centering
\includegraphics[scale=0.35]{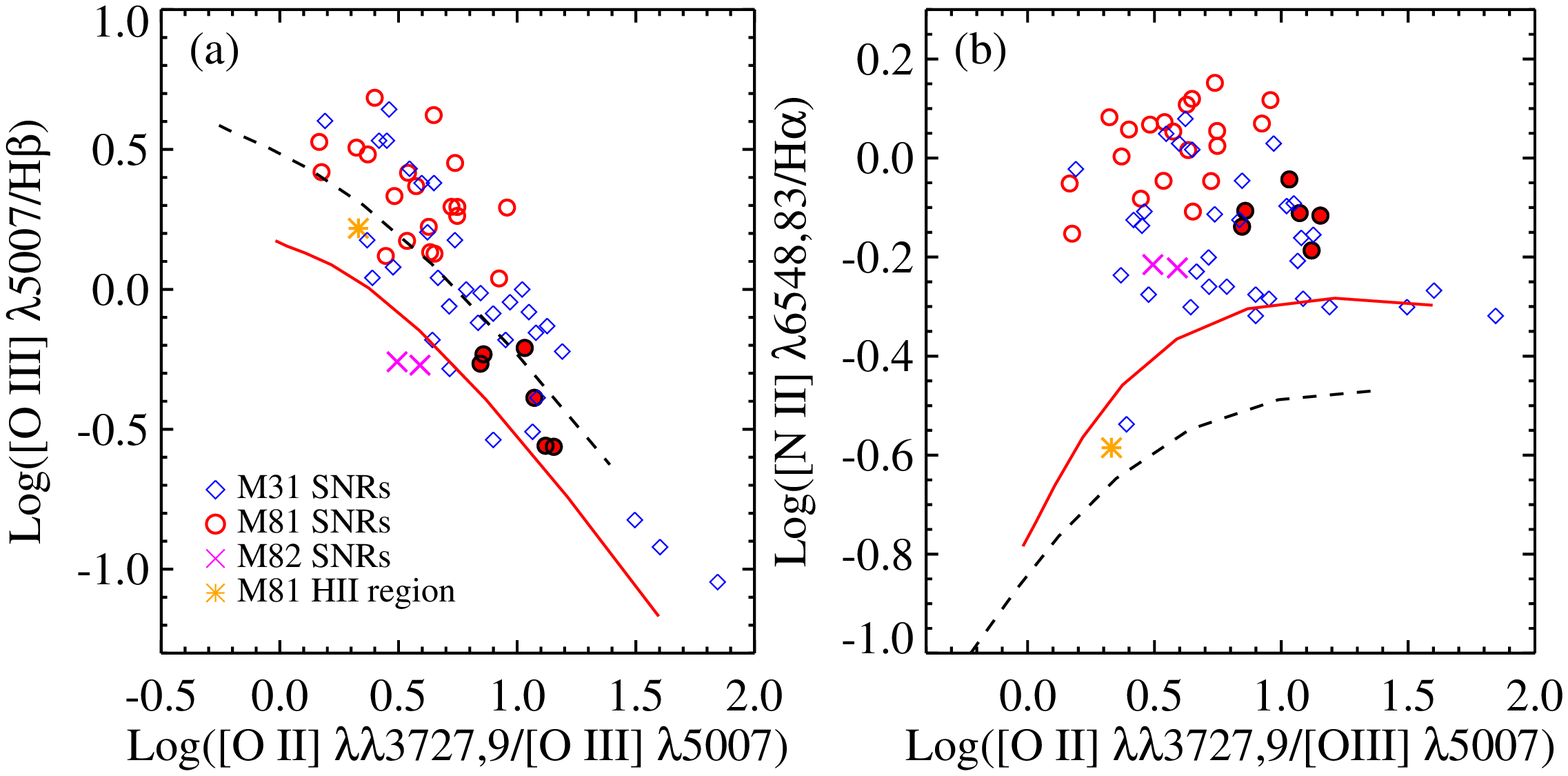}
\caption{
The \Othree/\Hb vs. \Otwo/\Othree~ diagram (a)
and the \Ntwo/Ha vs.  \Otwo/\Othree~ diagram (b)
of M81 SNRs (open circles), \Htwo region (asterisk), and M82 SNR candidates (crosses) in this study in comparison with M31 SNRs (diamonds, \citealp{gal99}). 
Open and filled circles denote \Othree-strong and \Othree-weak SNRs, respectively.
The dashed and solid lines represent  the upper limit of the theoretical models of \Htwo regions for one and two solar abundance  given by \citet{dop13}. }
\label{o23rat}
\end{figure}

{\bf Figure \ref{o23rat}(a)} displays the relation between \Othree $\lambda$5007/\Hb
 and \Otwo/\Othree $\lambda$5007 of the SNRs in M81 (this study) in comparison with those in M31 \citep{gal99}. 
We plotted also lines for
the theoretical models for \Htwo regions with one and two solar abundance given by \citet{dop13}. 
The following features are noted in this figure.
First, M81 SNRs are located clearly in two groups:
 an \Othree-weak group (\Othree $\lambda$5007/\Hb$<1$) and an \Othree-strong group (\Othree $\lambda$5007/\Hb$>1$). 
A majority of  M81 SNRs (20, 77 \%) belong to the \Othree-strong group, and 23\% (6) to the \Othree-weak group
 (L1, L2, L4, MF 22, MF29 and MF35).
 \citet{dop84} presented the strength of some emission lines as a function of shock velocity ($v_s$) in  the shock-ionization model for solar abundance in their Figures 5 and 6. According to these models, the value of the \Othree $\lambda$5007/\Hb ratio changes rapidly from --0.5 at $v_s \approx 80$ \kms~ to +0.5
at  $v_s \approx 100$ \kms, and increases slowly to $\approx 0.65$ at $v_s \approx 170$ \kms. 
Thus the \Othree $\lambda$5007/\Hb ratio becomes degenerate for $v_s<80$ \kms.
On the other hand, the \Otwo 3737/\Hb increases more slowly from 0.25 at $v_s=50$ \kms, to 1.1 at $v_s=90$ \kms~ so that it is a good indicator for shock velocity in the low shock velocity range.
The measured line ratios of the  \Othree-weak SNRs indicate that their shock velocity may be 65 to 80 \kms.
Second, the \Othree $\lambda$5007/\Hb ratio shows a strong correlation with the \Otwo $\lambda$3727/\Othree $\lambda$5007 ratio,
decreasing as  the \Otwo $\lambda$3727/\Othree $\lambda$5007 ratio increases.
M81 SNRs and M31 SNRs are located along the same sequence. However, the fraction of the \Othree-strong SNRs in M81 is larger than in M31. 
This indicates that the metallicity of M81 SNRs as well as M31 SNRs may be higher than the low metallicity adopted for the \Htwo region model.
Third, two M82 objects belong to the \Othree-weak group.

However, this is not the end of the story for the \Othree-weak SNRs. According to the modern shock models such as those in \citet{ho14}, the fast shocks would be \Othree-weak objects rather than \Othree-strong objects, and they are expected to have strong \Oone/\Hab, \Ntwo/Hab, and \Stwo/\Ha ratios. However, the \Othree-weak objects in Figures 11 and 13 show a large range of \Oone/\Hab, \Ntwo/Hab, and \Stwo/\Ha ratios. This suggests that these objects may be in part photo-ionized by stars, and are in part ionized by fast shocks with 200--300 \kms.

Similarly we plot the \Ntwo/\Ha versus \Otwo/\Othree$\lambda$5007 diagram of the SNRs  in {\bf Figure \ref{o23rat}(b)}.
It is found that the \Ntwo/\Ha ratio shows no correlation with the \Otwo/\Othree $\lambda$5007 ratio. This indicates that the \Ntwo/\Ha ratio depends little on shock conditions,
in contrast to the \Othree $\lambda$5007 ratio.
Also M81 SNRs  have, on average, higher \Ntwo/\Ha ratios
than M31 SNRs. 
This indicates that M81 SNRs may have higher nitrogen abundance compared with M31 SNRs.  
Two M82 SNRs are located in the region of M31 SNRs.
On the other hand, one M81 \Htwo region is far below the locations of the SNRs. 

\begin{figure}
\centering
\includegraphics[scale=0.35]{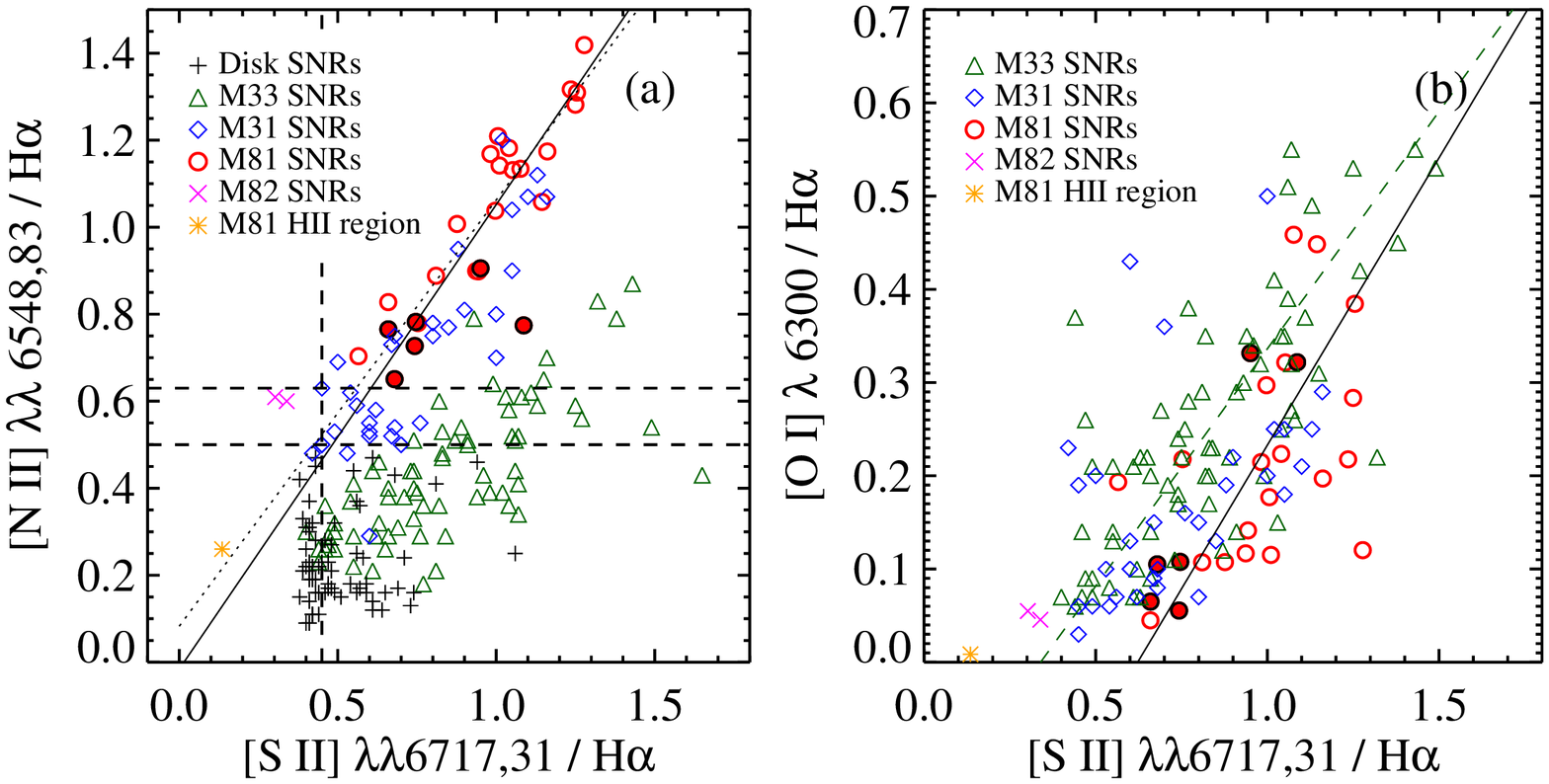}
\caption{
The \Ntwo/Ha vs. \Stwo/Ha diagram (a) and
the \Oone/\Ha vs. \Stwo/\Ha diagram (b) of
 M81 SNRs (circles), \Htwo region (asterisk), and M82 SNR candidates (crosses) in this study, in comparison with 
 M31 SNRs (diamonds, \citealp{gal99}), M33 SNRs (triangles, \citealp{gor98}), and SNRs in six disk galaxies (plusses, \citealp{leo13}).
Open and filled circles denote \Othree-strong and \Othree-weak SNRs, respectively.
Two dashed lines for \Stwo/\Ha=0.45 and \Ntwo/\Ha=0.5  were used for selecting M31 SNRs in \citet{gal99}, 
 and another for \Ntwo/\Ha=0.63 is a lower limit for shock ionization models.
In (a) the solid line represents a linear fit for M81 SNRs.
The dotted-line represents a linear fit result for M81 SNRs excluding one point (at \Stwo/\Ha$\sim 1.1$ and \Ntwo/\Ha$\sim 0.78$).
 In (b) the solid line and dashed line represent linear fits for M81 SNRs and M33 SNRs, respectively.}
\label{n2s2}
\end{figure}

{\bf Figure \ref{n2s2}(a)} displays the relation between \Ntwo/\Ha
 and \Stwo/\Ha ratios of the SNRs in M81 (this study) as well as those
 in M31, M33,  and other disk galaxies \citep{gal99, gor98, leo13}. 
We plotted also the boundary lines at \Stwo/\Ha=0.45 and \Ntwo/\Ha=0.5, 
 which were used for selecting M31 SNRs by \citet{gal99},
 and a line at \Ntwo/\Ha$=0.63$ (Log \Ntwo/\Ha$=-0.2$), a lower limit for shock ionization models.
Several features are noted in this figure.
First, all M81 SNRs are located above the lower limit for shock ionization models, 
 while some of the SNRs in other galaxies are below this limit.
Second, M81 SNRs show a remarkably strong correlation between these ratios, which is fitted linearly well with a slope close to one:
 \Ntwo/\Ha = ($0.934\pm0.080$) \Stwo/\Ha $+ (0.112\pm0.071)$ with $rms=0.109$
 (\Ntwo/\Ha = ($0.979\pm0.066$) \Stwo/\Ha $+ (0.083\pm0.065)$ with $rms=0.081$, if one outlier is excluded).  This strong correlation shows that \Ntwo/\Ha can be also useful to selecting SNRs, when \Stwo~ lines are not available.
Third, the SNRs in other galaxies also show similar correlatons, but they are all located below the sequence for M81 SNRs, with larger scatters.

In the study of M33 SNRs, \citet{gor98} noted that \Oone/\Ha ratios show a strong correlation with \Stwo/Ha ratios.
In {\bf Figure \ref{n2s2}(b)} 
we display the relation between the \Oone $\lambda$6300/\Ha
 and \Stwo/\Ha ratios of the SNRs in M81 in comparison with M31 SNRs and M33 SNRs.
M81 SNRs also show a reasonable correlation between these ratios.  
M31 SNRs also show a similar correlation, if seven SNRs with high \Oone $\lambda$6300/\Ha ratios are excluded.
From the linear fits  we derive 
\Oone $\lambda$6300/\Ha = $( 0.616 \pm 0.080 )$ \Stwo/\Ha $- (0.384 \pm 0.075)$ with $rms = 0.115$ for M81, and  
\Oone$\lambda$ 6300/\Ha = $( 0.664 \pm 0.065 )$ \Stwo/\Ha $- (0.301 \pm 0.053)$ with $rms = 0.117$ for M33.
The slope of this relation is much smaller than the value for the \Ntwo/\Ha and
\Stwo/\Ha relations. 
In addition, M81 SNRs are located below the M33 and M31 sequences, which is the opposite to the case of the \Ntwo~ ratios.
\Oone $\lambda$6300/\Ha ratios can be also used for SNR selection, when \Stwo~ lines are not available.

The tight correlations between \Ntwo, \Stwo, and \Oone~ and little correlation between these lines and \Othree~ lines found in this study are consistent with the results of the modern shock models (Dopita, private communication). Faster shocks increase the internal UV radiation field, making the recombination zone more extensive. Therefore, as the shock velocity increases, the line ratios of \Ntwo/\Hab, \Stwo/\Hab, and \Oone/\Ha increase,  while \Othree/\Hb decreases. The \Othree~ line emitting region is not in the recombination zone, so is unaffected, while the \Hb line that is emitted from the recombination zone becomes stronger, decreasing the \Othree/\Hb ratio.

\begin{figure}
\centering
\includegraphics[scale=0.5]{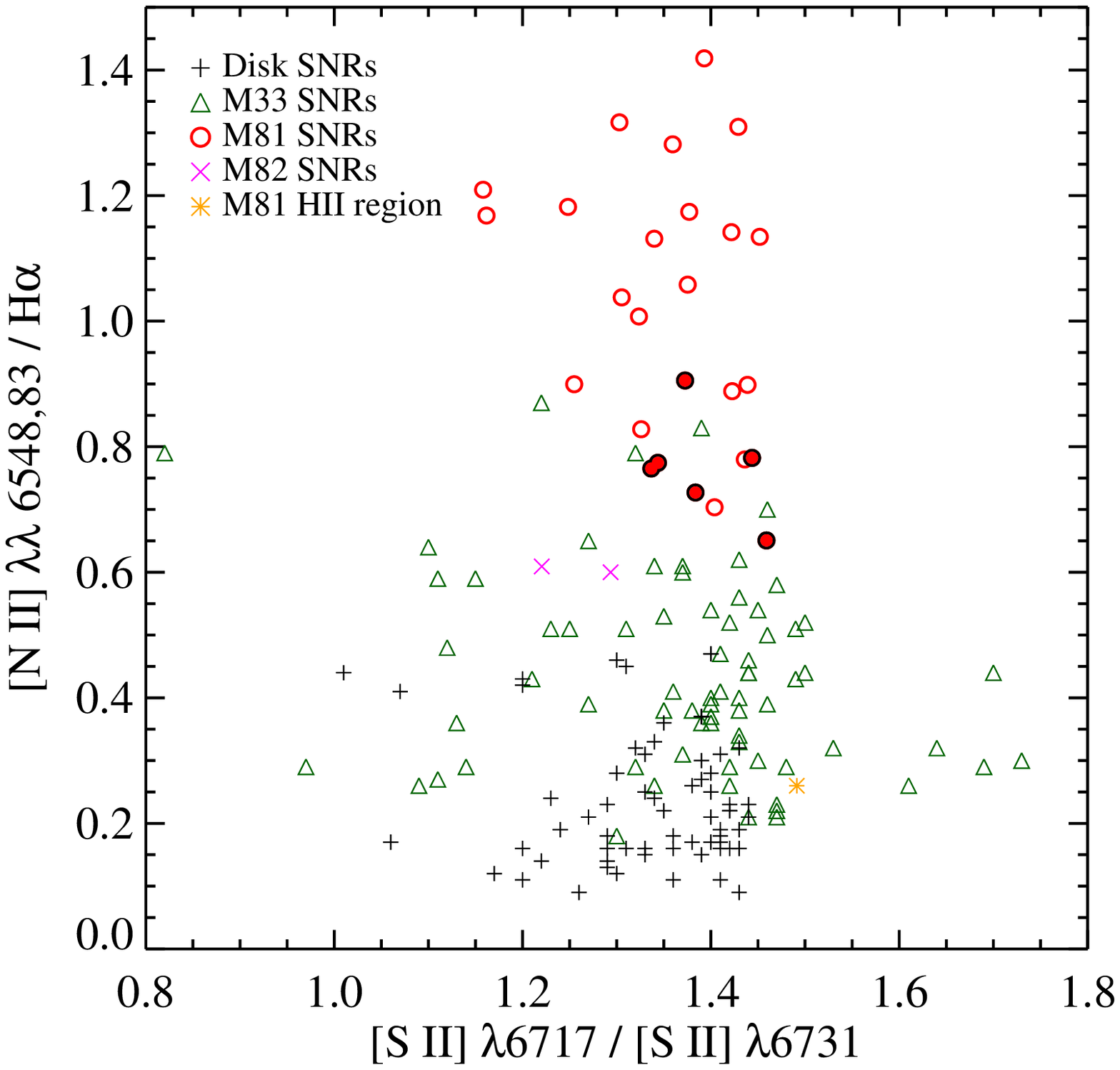}
\caption{
The \Ntwo/\Ha vs. \Stwo$\lambda$6717/\Stwo$\lambda$6731 diagram of 
 M81 SNRs (circles), \HtwoR~ (asterisk), and M82 SNR candidates (crosses) in this study, in comparison with 
 M31 SNRs (diamonds, \citealp{gal99}), M33 SNRs (triangles, \citealp{gor98}), and SNRs in six disk galaxies (plusses, \citealp{leo13}).
Open and filled circles denote \Othree-strong and \Othree-weak SNRs, respectively. }
\label{n2s2rat}
\end{figure}

{\bf Figure \ref{n2s2rat}} 
 displays the \Ntwo/\Ha ratios against 
 the \Stwo$\lambda$6717/\Stwo$\lambda$6731 ratios that are an density indicator,
 for the SNRs in M81 (this study) as well as 
 in M31, M33, and other disk galaxies \citep{gal99, gor98, leo13}. 
M81 SNRs show little correlation between these two ratios, as those in other galaxies. 
This shows that \Ntwo/\Ha ratios depend little on the density of the SNRs.

\subsection{Comparison with AGN Classification Diagrams}
Emission line ratio diagrams are often used  for spectral classification of AGNs as well as SNRs. 
\citet{bal81} suggested classification parameters for the emission-line spectra of extragalactic objects,
 arguing that the \Othree $\lambda$3727/\Othree $\lambda$5007 ratio is efficient to distinguish photo-ionized objects (\Othree $\lambda$3727 $<$ \Othree $\lambda$5007)
 and shock-heated objects (\Othree $\lambda$3727 $>$ \Othree $\lambda$5007) (called BPT diagrams). 
However, this parameter is affected much by extinction. Later \citet{vei87}
 suggested other parameters that depend much less on extinction:
 \Othree $\lambda$5007/\Hb versus \Ntwo$\lambda$ 6583/\Hab,
 \Othree $\lambda$5007/\Hb versus \Stwo $\lambda\lambda$6717,31/\Hab, and
 \Othree $\lambda$5007/\Hb versus \Oone $\lambda$6300/\Ha (called VO diagrams). 
Classification boundaries on these diagrams were given for \Htwo regions, Seyfert galaxies, and LINERs by \citet{shi90}.
Later \citet{kew01} presented a boundary for the starburst galaxies and AGN based on theoretical models, 
 which was revised for pure star-forming galaxies in \citet{kau03}.
These diagrams are useful also for distinguishing SNRs and \Htwo regions.

\begin{figure}
\centering
\includegraphics[scale=0.45]{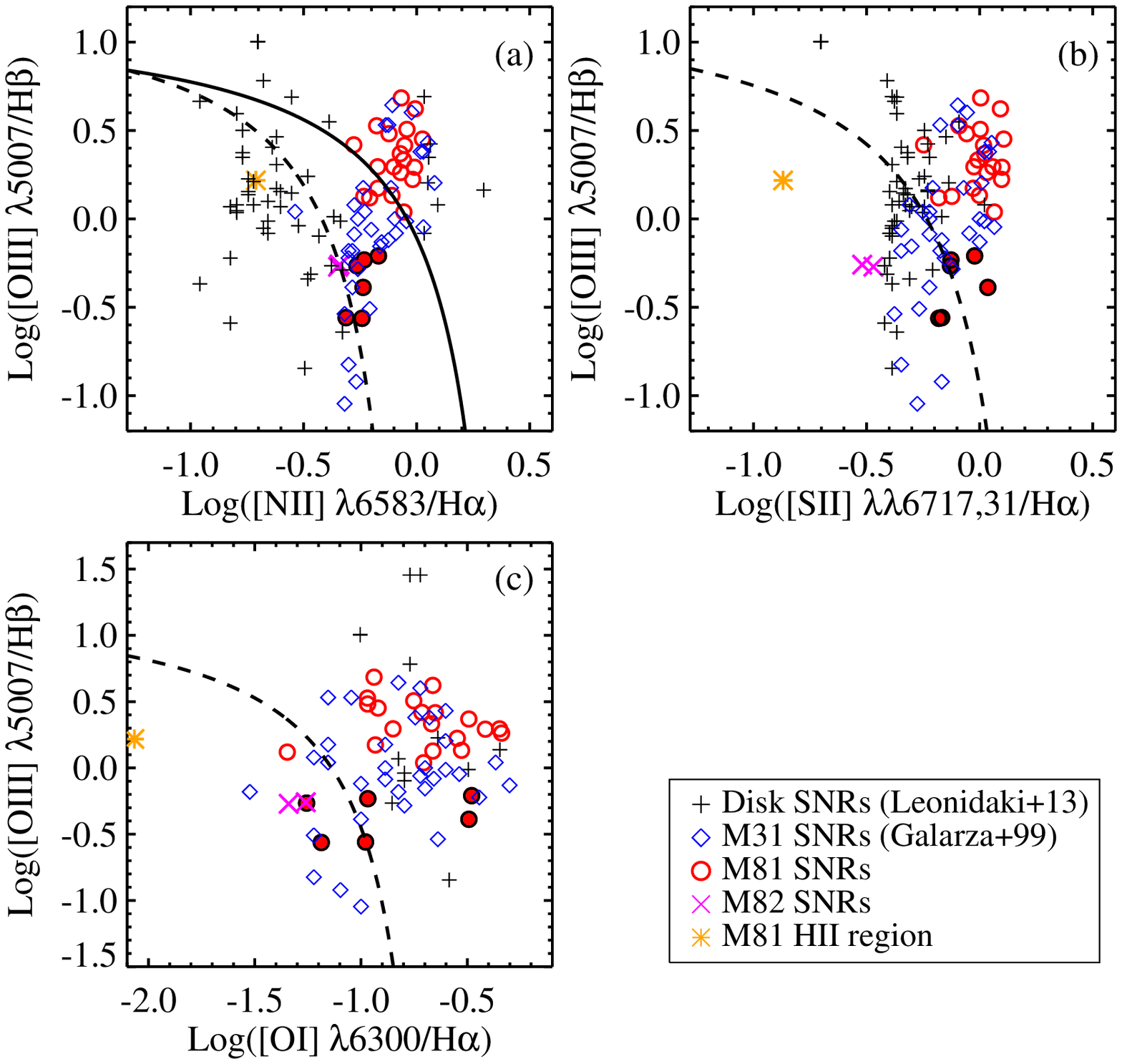}
\caption{Spectral classification diagrams (BPT or VO diagrams) in the logarithmic scales of line ratios of the SNRs in M81 (circles, this study) 
 in comparison with those in M31 (diamonds, \citealp{gal99}) and other disk galaxies (plusses, \citealp{leo13}):
(a) \Othree $\lambda$5007/\Hb vs. \Ntwo $\lambda$6583/\Hab,
(b) \Othree $\lambda$5007/\Hb vs. \Stwo $\lambda\lambda$6717,31/\Hab, and
(c) \Othree $\lambda$5007/\Hb vs. \Oone $\lambda$6300/\Hab.
The curved solid line and dashed line denote a theoretical upper limit for starburst galaxies \citep{kew01}, and a boundary for pure star-forming galaxies \citep{kau03}.  
Note that all 26 targets in M81 are located in the SNR region and 
 that M81 SNRs are located in two groups: 
 an \Othree-strong group (open circles) and an \Othree-weak group (filled circles).
Two M82 objects are located around the boundary between \Htwo regions and SNRs, while one M81 \Htwo region is inside the \Htwo region boundary. }
\label{bpt}
\end{figure}

{\bf Figure \ref{bpt}} displays these diagrams for M81 SNRs in this study.
We also plotted the data for the SNRs in M31 and other disk galaxies \citep{gal99,leo13} for comparison.
Most of the SNRs are located above
 the Kauffmann's demarcation line in Figure \ref{bpt}(a), and all of them are outside the \Htwo region boundaries, 
 while one \Htwo region is located inside the \Htwo region boundary.
Most of M81 SNRs are located in the LINER region, while a few are in the Seyfert region.
This shows again that M81 SNRs are shock-ionized.

\begin{figure}
\centering
\includegraphics[scale=0.4]{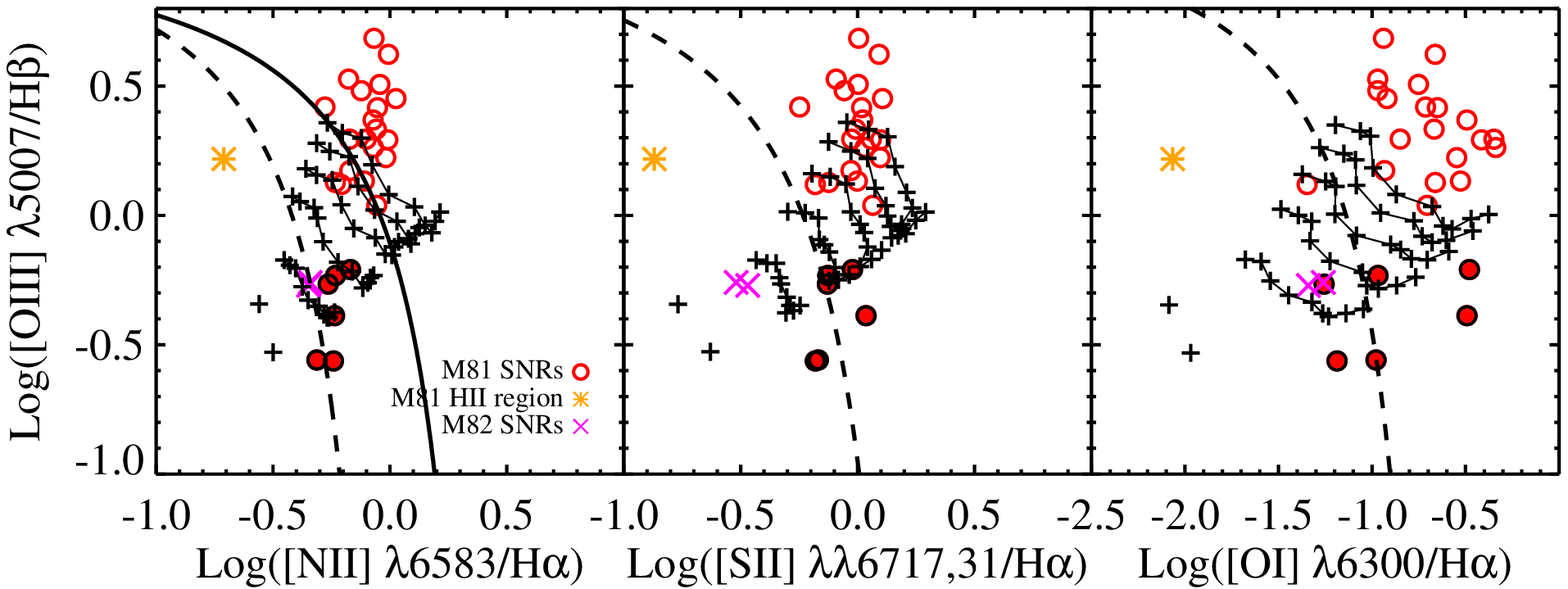}
\caption{Same as Fig. 13, but for shock ionization models  (crosses), given in \citet{ho14}, for 12+Log(O/H)=9.14, shock fraction from 0 to 100 \% with 20 \% interval (from bottom to top), and shock velocity from 100 to 300 \kms~ (with 20 \kms~ interval for shock velocity 100 to 200 \kms, and 25 \kms~ interval for shock velocity 200 to 300 \kms) (from left to right). \\}
\label{BPTHo}
\end{figure}

It is noted that the two M82 objects are located close to the boundary of the star-forming regions in the \Othree $\lambda$5007/\Hb versus \Ntwo$\lambda$ 6583/\Ha diagram, 
and are inside the boundary  in the
\Othree $\lambda$5007/\Hb versus \Stwo $\lambda\lambda$6717,31/\Hab, and
 \Othree $\lambda$5007/\Hb versus \Oone $\lambda$6300/\Ha, as some of the \Othree-weak objects are. 
In Figure \ref{BPTHo}, we plotted the same diagrams as Figure \ref{bpt}, overlaying shock ionization models that \citet{ho14} presented for 12+Log(O/H)=9.14, shock fraction from 0 to 100\%, and shock velocity from 100 to 300 \kms. \citet{ho14} used these models to explain the shock features in the outflow region of star-forming galaxies in terms of combination of shock and photo-ionization. These models were given for a fixed oxygen abundance value, but they are still useful as a reference to explain the data for M82.  
The data of the M82 objects are consistent with the fast shock models with a low fraction of shock in the figure. This and their location in the galactic outflow region indicate that the two M82 objects may be shock condensations in the general outflow perpendicular to the major axis of M82. However, the shell structures seen in Figure 4 indicate that they may be SNRs as well.

\subsection{Comparison with Shock-ionization Models}

\citet{dop84} presented the shock-ionization models for weak shock velocity ($v_s <200$ \kms). 
Later these models were extended to the case of fast shock models covering the shock velocity of 200 to 1000 \kms \citep{dop95,dop96,all08}.
We compared the line ratios of M81 SNRs with the models given by \citet{dop84} that are useful for the analysis of SNRs in nearby galaxies.
These models were applied previously to the case of M31 and M33 SNRs by \citet{bla85} and \citet{smi93}.
However, the models of \citet{dop84} are based on old atomic physics, which is much improved nowadays.
Emission line ratios of shock models depend not only on abundances of the interstellar medium, but also on the complexity of partially radiative shocks, degree of mixing with photoionization, shock velocity, the strength of magnetic field, and grain destruction. 
However, these factors were not included in the old models, and no modern shock model grids considering all these factors are not yet available. 
Therefore the results of determination of abundances of the SNRs based on the old simple models should be considered only as an approximate guide.
These results can be improved, when the new modern shock models are available in the future.
 
\begin{figure}
\centering
\includegraphics[scale=0.35]{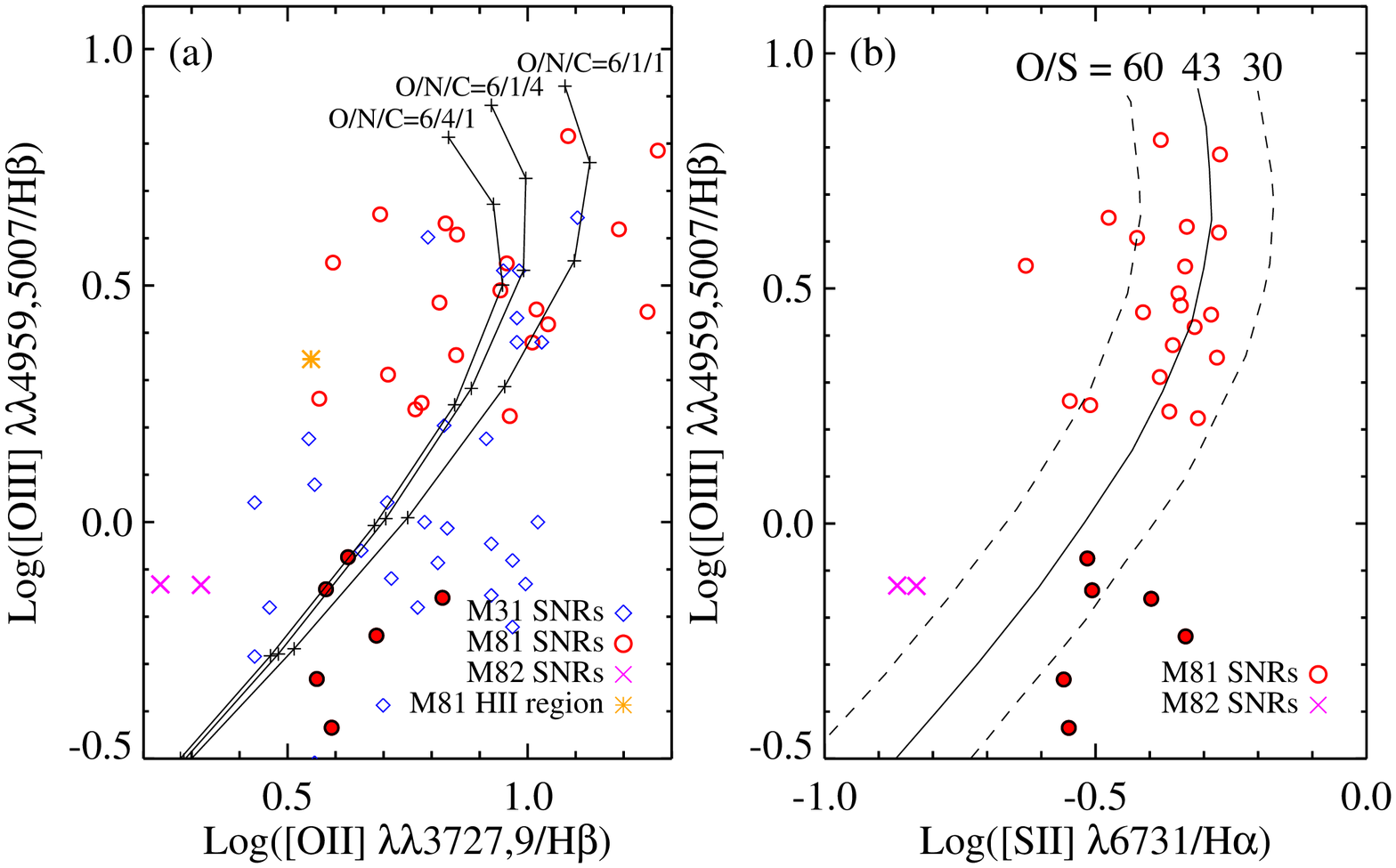} 
\caption{The \Othree $\lambda\lambda$4959,5007/\Hb vs. \Otwo $\lambda\lambda$3727,9/\Hb diagram (a) and
the \Othree $\lambda\lambda$4959,5007/\Hb vs. \Stwo $\lambda$6731/\Ha diagram (b) of M81 SNRs (circles) and HII region (cross) 
in comparison with
the shock ionization model grid for shock velocity $v_s = 106$ \kms~  
and varying ratios of oxygen, nitrogen, carbon, and sulfur abundances (lines) provided by \citet{dop84} (their Figures 7 and 9).
Open and filled circles denote \Othree-strong and \Othree-weak SNRs, respectively. }
\label{o3o2grid}
\end{figure}

We plot in {\bf Figure \ref{o3o2grid}(a)} the \Othree/\Hb versus \Otwo/\Hb diagram of M81 SNRs derived in this study and  M31 SNRs \citep{gal99} in comparison with the shock-ionization models for the shock velocity of $v_s =106$ \kms~ and the fixed abundance ratio of $O/S=42.8$ given by \citet{dop84}.
It is noted that the SNRs in M81 and M31 are located 
roughly around the model grids, but their scatter in \Otwo\Hb ratios is much larger than the range of the grids.
This indicates again that the large scatter in \Otwo\Hb ratios  is mainly due to the large range in shock velocity of the SNRs. 
It is noted that the large scatter in this figure can also be produced if the shocks are too young to be fully radiative, being in partially radiative state. According to the radiative shock models applied to the case of the microquasar S26 in NGC 7793 by \citet{dop12}, the \Othree/\Hb ratio decreases, while the \Otwo/\Hb ratio increases, as shocks get older.
Therefore the large scatter in the \Othree/\Hb versus \Otwo/\Hb diagram must be due to the combination of a large range of shock velocity and shock age.

In {\bf Figure \ref{o3o2grid}(b)} 
 we show the \Othree/\Hb versus \Stwo $\lambda$6731/\Ha diagram of M81 SNRs (circles) and M82 SNRs (crosses) in comparison with the same shock ionization models but for varying $O/S$ abundance ratios.    
It is seen that  the \Othree-strong SNRs are located mostly around the grid for O/S$=43$, which is similar to the mean value adopted for other grids, O/S$=42.8$ by \citet{dop84}.

\begin{figure}
\centering
\includegraphics[scale=0.35]{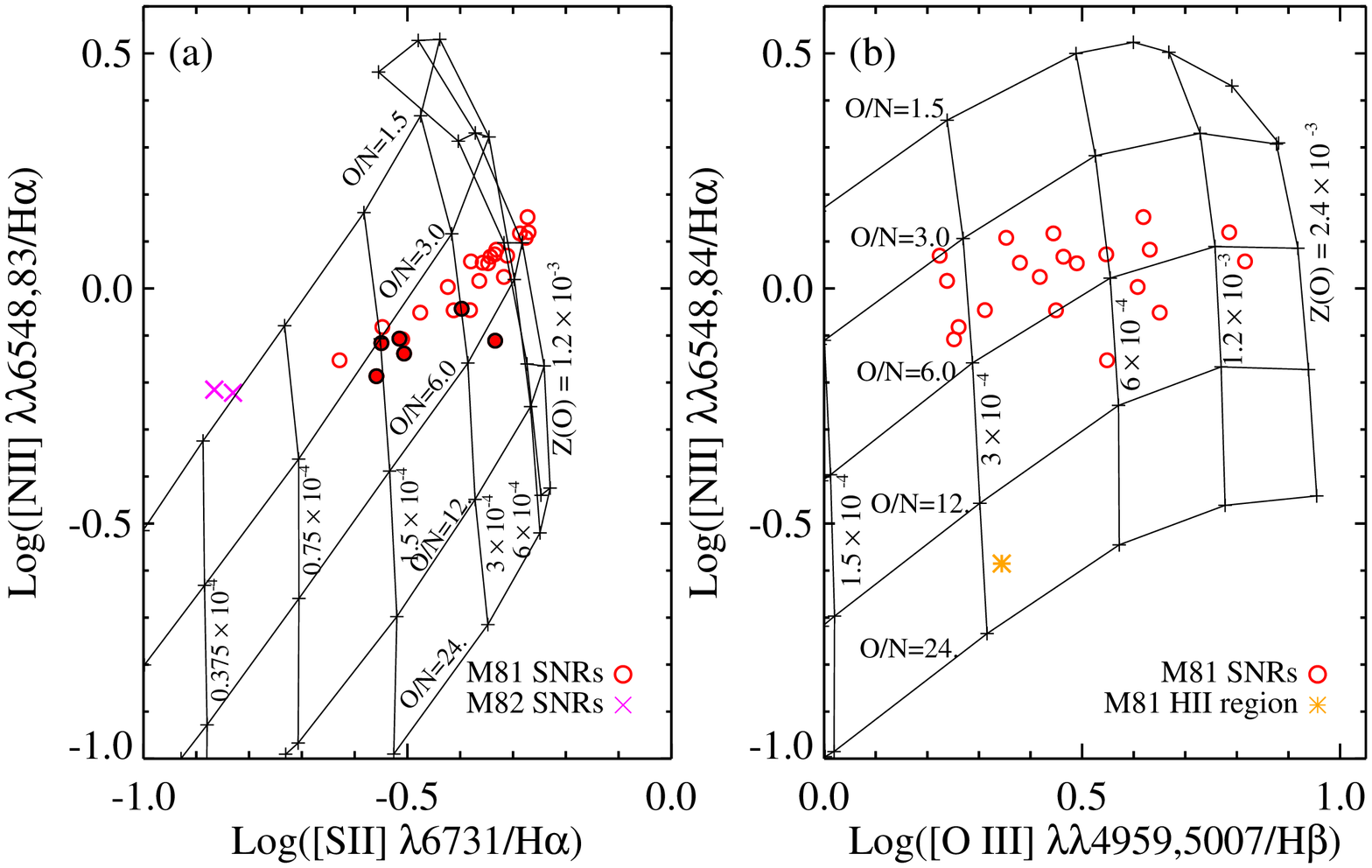}
\caption{
The \Ntwo $\lambda\lambda$6548,83/\Ha vs. \Stwo $\lambda$6731/\Ha  
diagram (a) and the \Ntwo $\lambda\lambda$6548,83/\Ha vs. \Othree $\lambda\lambda$4959,5007/\Hb diagram (b) 
of M81 SNRs (circles) and M82 SNRs (crosses) in comparison with the shock ionization model grid   
for O/S=42.8 by \citet{dop84} (their Figures 8 and 10). 
Open and filled circles denote \Othree-strong and \Othree-weak SNRs, respectively. 
Note that \Othree-weak SNRs are beyond the left limit in (b).}
\label{n2s22grid}
\end{figure}

{\bf Figure \ref{n2s22grid}(a)} displays the \Ntwo/\Ha versus \Stwo $\lambda$6731/\Ha diagram of M81 SNRs 
 in comparison with the same shock-ionization models for various values of oxygen abundance and the ratio of oxygen to nitrogen abundance (for the abundance ratio of O/S$=42.8$). 
\Stwo $\lambda$6731/\Ha ratios are sensitive to the ratio of oxygen and sulfur abundance,  but loses its sensitivity at the high abundance.
\Ntwo/\Ha ratios are a good indicator for nitrogen abundance.
Some of M81 SNRs are located beyond the model grid limits, 
 for which abundances cannot be derived.
Note also that the \Ntwo/\Ha versus \Stwo/\Ha can be a good calibrator of the N/S abundance ratio, since all of these lines arise in the same region of the shock.

In {\bf Figure \ref{n2s22grid}(b)}
we show the \Ntwo/\Ha versus \Othree/\Ha diagram of M81 SNRs
 in comparison with the same shock-ionization models.
In this figure, \Othree/\Ha ratios appear to be a good indicator of oxygen abundance.
The \Othree-strong SNRs in M81 are located inside the model grid,
 while the \Othree-weak SNRs are beyond the lower limit of the model grid. These \Othree-weak SNRs might have low abundance or low shock velocity, or both.
Thus the \Othree/\Ha ratio is a better abundance indicator than the \Stwo $\lambda$6731/\Ha,  but not as good as claimed by \citet{dop84}.
We derived nitrogen and oxygen abundances of M81 SNRs using the model grids in {\bf Figure \ref{n2s22grid}}, listing them in Table \ref{ratio}.

\begin{figure}
\centering
\includegraphics[scale=0.5]{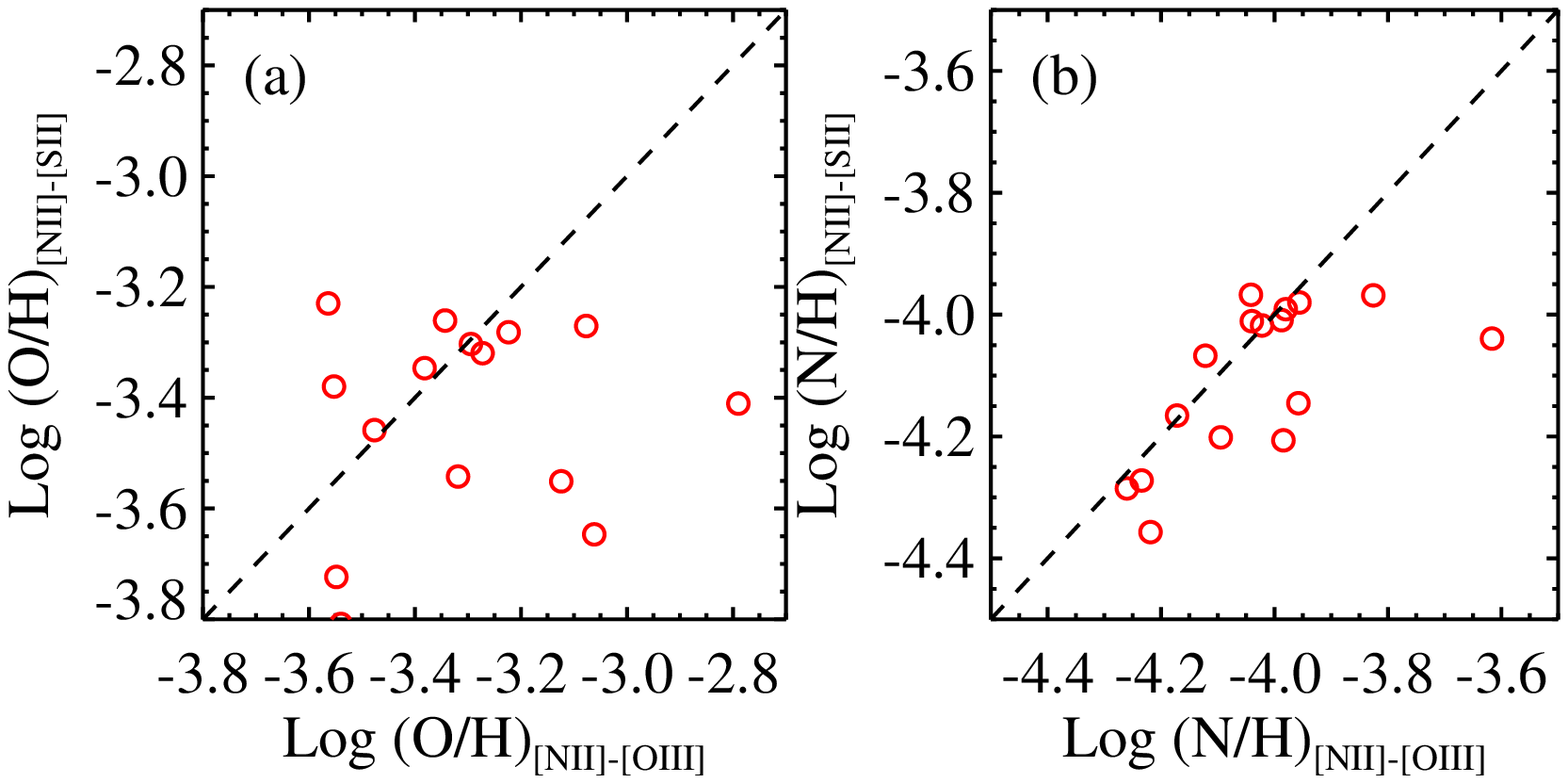}
\caption{Comparison of the oxygen (left) and nitrogen (right) abundances of M81 SNRs derived from the \Ntwo/\Ha vs. \Stwo$\lambda$6731/\Ha grid and the \Ntwo/\Ha vs. \Othree/\Ha grid.
Dashed lines represent one-to-one relations. \\}
\label{zOzNcomp}
\end{figure}

In {\bf Figure \ref{zOzNcomp}} we compare the oxygen and nitrogen abundances derived using the two diagrams. 
O/H values show a large scatter and little correlations between two estimates.
On the other hand, N/H values show a correlation between the two estimates so that 
the estimated values of N/H are considered to be reliable.
We conclude that the \Ntwo/\Ha versus \Othree/\Ha diagram is better for abundance estimation of the SNRs than the \Ntwo/\Ha versus \Stwo $\lambda$6731/\Ha diagram.  
However, it is noted that  \Othree/\Ha ratios are more affected by shock velocities
 so that we need to be cautious in using the oxygen abundances derived from \Othree/\Ha ratios.

We calculated also the values of the metallicity index A defined by \citet{dop84}, 
 Log A = Log O/H + Log N/H + Log S/H, for the fixed value of O/S$=42.8$.
 These values are listed in Table \ref{ratio}.
The metallicity index A can be a good metallicity indicator, if
reliable values of oxygen, nitrogen, and sulfur abundance can be
derived from the data of SNRs. However, oxygen lines can be affected significantly by shock conditions, while nitrogen lines depend mainly on abundance. From this we consider that the metallicity index A derived in this study is not a good indicator of metallicity for SNRs. Therefore we do not discuss the values of A derived in this study, although we list them in the table for future studies.

\subsection{Relations between Line Ratios and Sizes of SNRs}

\begin{figure}
\centering
\includegraphics[scale=0.5]{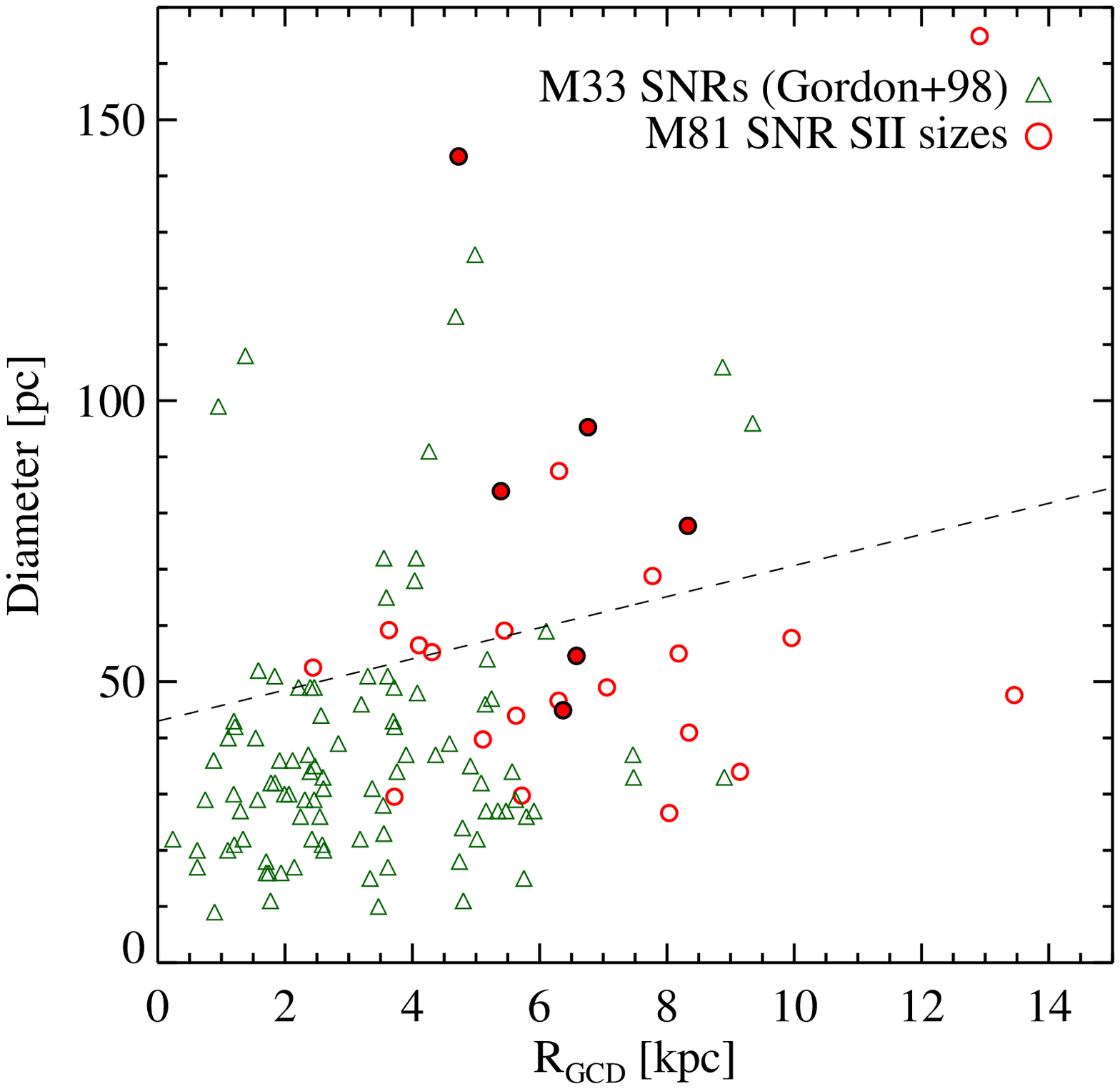}
\caption{The size vs. galactocentric distance of M81 SNRs (circles) in this study in comparison with M33 SNRs (triangles, \citealp{gor98}). 
Open and filled circles denote \Othree-strong and \Othree-weak SNRs, respectively. 
Dashed line represent a linear fit for the sample of all SNRs in M81, with a slope of $2.97\pm3.58$. 
It appears to show a weak correlation between
the size and galactocentric distance, but the correlation coefficient is as small as 0.22, showing little correlation. } 
\label{sizeR}
\end{figure}

In {\bf Figure \ref{sizeR}} we plot the diameter versus galactocentric distance of M81 SNRs (this study) in comparison with M33 SNRs \citep{gor98}. 
It is noted that four of six large SNRs with $D>70$ kpc are \Othree-weak SNRs. 
It appears that M81 SNRs show a weak  correlation between the size and the galactocentric distance. 
However, a linear fit for the sample of all SNRs in M81 yields a slope of $dD/dR = 2.97\pm3.58$, and the correlation coefficient is as small as 0.22, showing little correlation between the two parameters. 
Similarly M33 SNRs show little correlation between the two parameters \citep{gor98}. 
It is noted that only small SNRs with $D<60$ pc are seen in the inner region at $R<4.5$ kpc, while a large range of SNRs are seen in the outer region at $R>4.5$ kpc. This trend is also found for the SNRs in M31 and M33 \citep{lee14a,lee14b}. This trend is consistent with the explanation that the pressure in the interstellar medium is higher closer to the galaxy center, and the SNRs become radiative more quickly in the inner region by consequence so that they evolve faster and die away \citep{dop10}.

\begin{figure}
\centering
\includegraphics[scale=0.5]{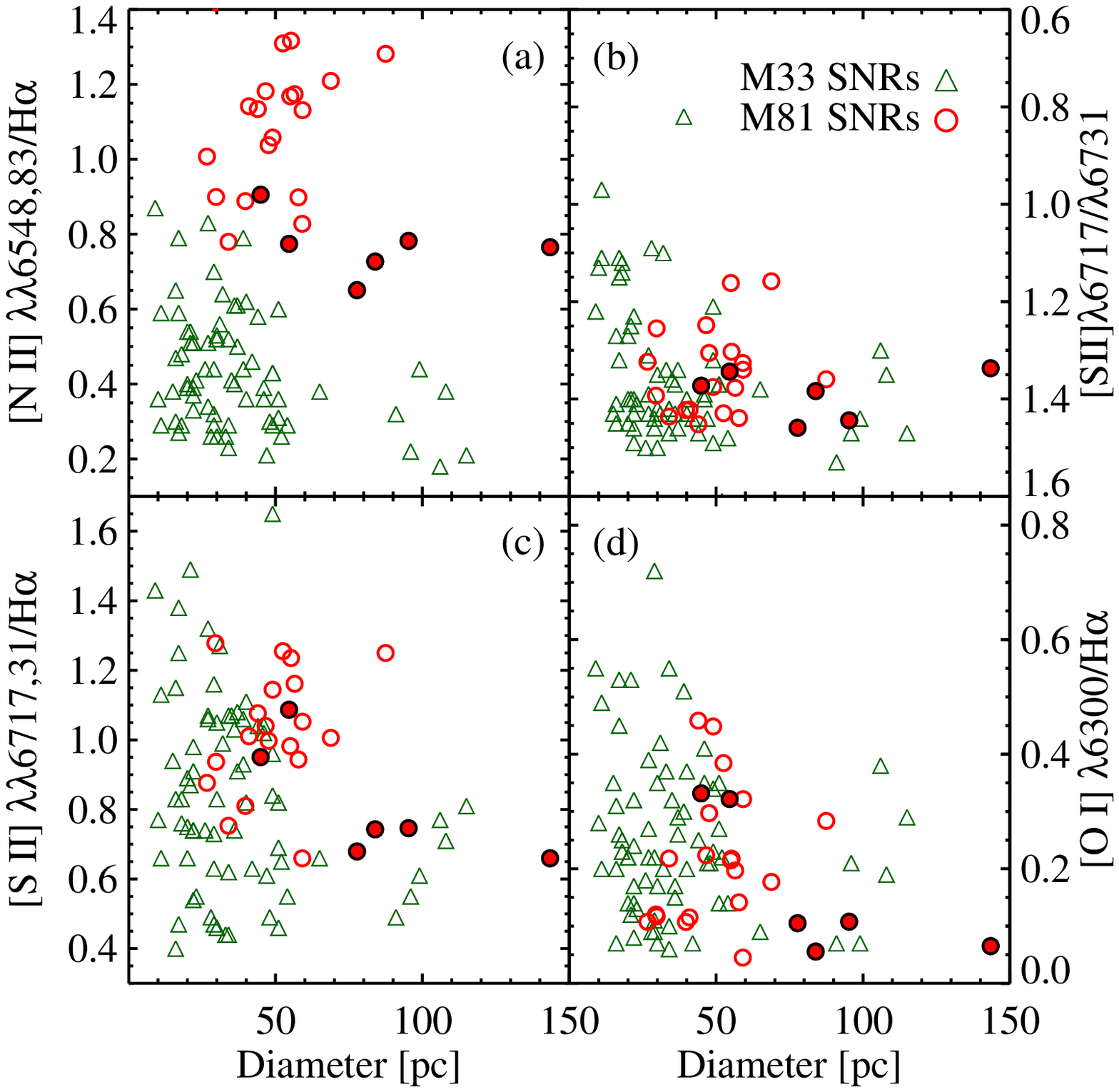}
\caption{Emission line ratios vs. size of  M81 SNRs  (circles) in this study in comparison with M33 SNRs  (triangles, \citealp{gor98}): 
(a) \Ntwo/\Ha, (b) \Stwo $\lambda$6717/\Stwo $\lambda$6731,
(c) \Stwo/Ha, and (d) \Oone $\lambda$6300/Ha.
Open and filled circles denote \Othree-strong and \Othree-weak SNRs, respectively. }
\label{linesize}
\end{figure}

{\bf Figure \ref{linesize}} displays the relations of four line ratios (\Ntwo/\Hab, \Stwo $\lambda$6717/\Stwo $\lambda$6731, \Stwo/\Hab, and \Oone/\Hab) and
 sizes of  M81 SNRs (this study) in comparison with M33 SNRs \citep{gor98}.
This figure shows a few notable features.
First, small SNRs with $D<70$ pc in M81 show a larger scatter in the ratios of these lines. 
Second, large SNRs with $D>70$ pc are mostly (four out of six) \Othree-weak SNRs. 
The \Othree-weak SNRs show lower values with a much smaller scatter in the ratios of these lines. 
However, two large SNRs (MF 06, MF 17) have high values. 
Considering that the \Othree-weak SNRs have fast shock velocity and they are relatively large, we conclude that they may be dynamically old.
Third, small and large SNRs in M33 show similar behaviors. 
However, M81 SNRs have much higher \Ntwo/\Ha ratios than M33 SNRs.
Fourth, the \Stwo $\lambda$6717/\Stwo $\lambda$6731 ratios of small SNRs in M81 (as well as in M33) are
 on average larger than those of the large SNRs, 
 showing that larger SNRs have in general lower densities than smaller SNRs, as expected.

\subsection{Radial Variation of Sizes, Line Ratios, and Abundances of SNRs}

\begin{figure}
\centering
\includegraphics[scale=0.5]{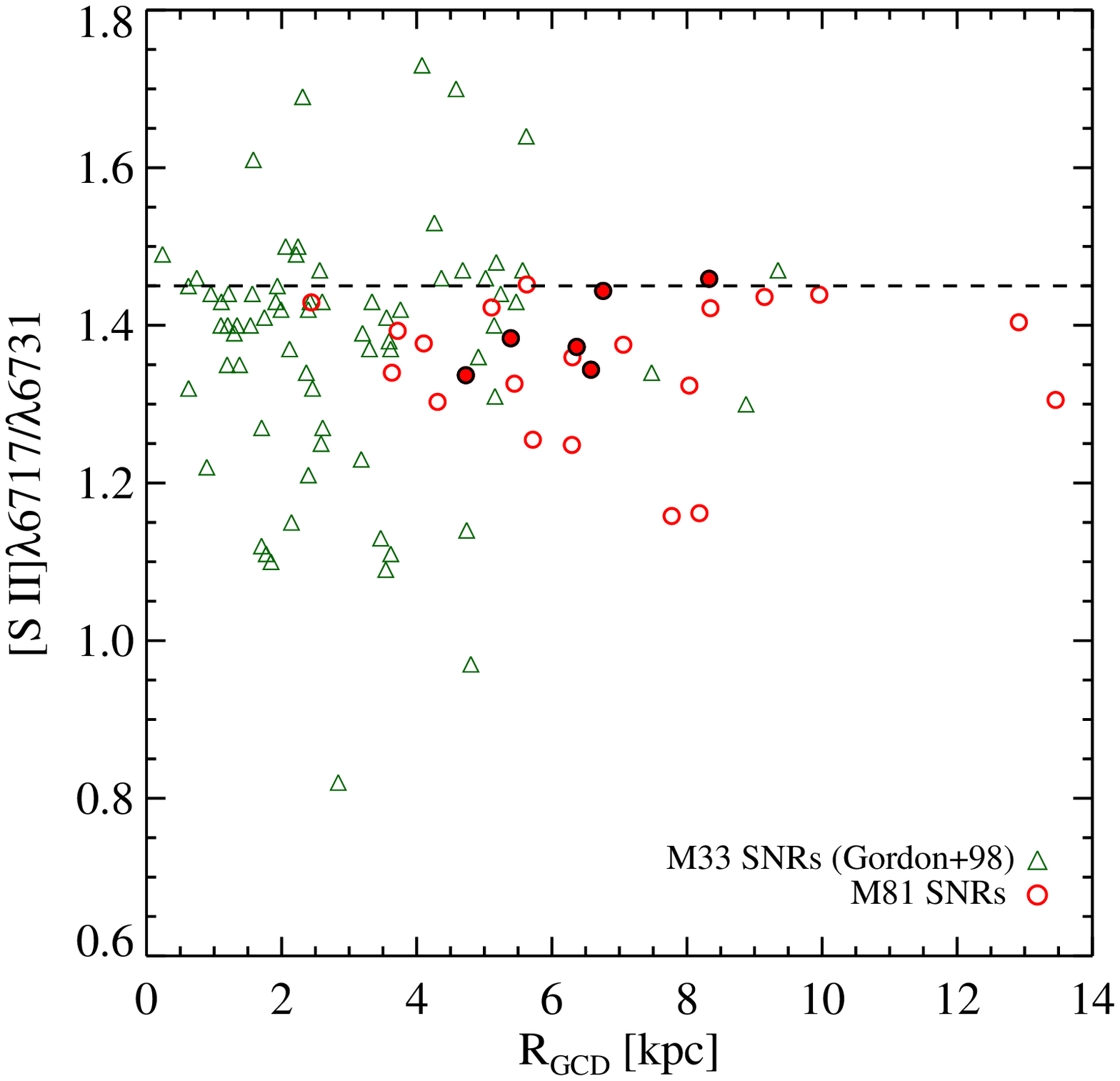}
\caption{
The \Stwo $\lambda$6717/\Stwo $\lambda$6731 ratio vs. galactocentric distance of M81 SNRs (circles) in this study with M33 SNRs (triangles, \citealp{gor98}).  
Open and filled circles denote \Othree-strong and \Othree-weak SNRs,
respectively. } 
\label{s2ratR}
\end{figure}

{\bf Figure \ref{s2ratR}} plots the \Stwo $\lambda$6717/\Stwo $\lambda$6731 ratio versus galactocentric distance of M81 SNRs as well M33 SNRs \citep{gor98}.
What is the most impressive about this figure is that M81 SNRs show clearly an upper envelope at \Stwo $\lambda$6717/\Stwo $\lambda$6731 $\approx 1.45$.
In comparison, M33 SNRs show a larger scatter, and
 some of them have \Stwo $\lambda$6717/\Stwo $\lambda$6731 values above the envelope.
The latter might be due to the larger errors in the measurements of M33 SNRs \citep{gor98}.
The value for the upper envelope of M81 SNRs is close to a theoretical lower limit, 1.43, \citep{bla85}. This shows that the measurements of \Stwo $\lambda$6717/\Stwo $\lambda$6731  ratios of M81 SNRs in this study are solid. 
All M81 SNRs have \Stwo $\lambda$6717/\Stwo $\lambda$6731 $> 1.15$.
This value corresponds to the density of 312 cm$^{-3}$, according to the conversion of \Stwo $\lambda$6717/\Stwo $\lambda$6731 ratio into densities 
 as a function of temperature in Figure 7 of  \citet{bla85}.
This shows that they are located in the low density region.
M81 SNRs show little radial gradient in this line ratio.

\begin{figure}
\centering
\includegraphics[scale=0.45]{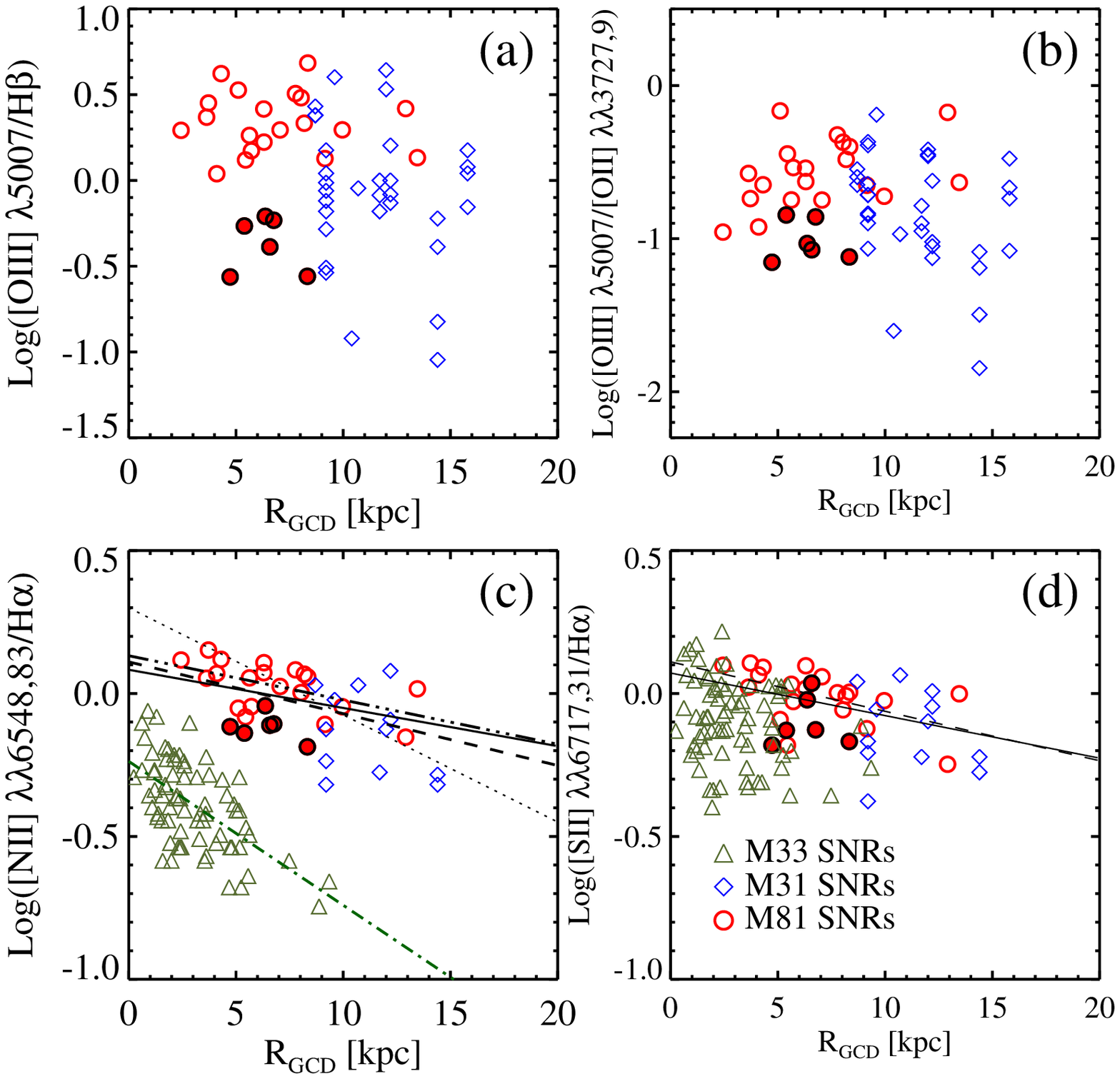}
\caption{Several line ratios vs. galactocentric distance for M81 SNRs 
in comparison with M31 SNRs (diamonds, \citealp{gal99}) and M33 SNRs (triangles, \citealp{gor98}): 
(a) Log \Othree/\Hab, (b) Log \Othree/\Otwo,  (c) Log \Ntwo/\Ha, and (d) Log \Stwo/\Hab.
Open and filled circles denote \Othree-strong and \Othree-weak SNRs, respectively. 
The solid line and dashed line in (c) represent linear fits for all M81 SNRs and for M81 SNRs excluding one outlier at about 12 kpc, with slopes,
 $-0.013\pm0.007$ and $-0.018\pm0.008$ dex kpc$^{-1}$, respectively.
The dotted-line is a linear fit for M31 SNRs given by \citet{gal99} with a slope  $-0.04$ dex kpc$^{-1}$, and the dot-dashed line a linear for M33 SNRs,
with a slope  $-0.05$ dex kpc$^{-1}$.
The solid-line in (d) represents a linear fit for all M81 SNRs with a slope, $-0.015\pm0.008$ dex kpc$^{-}$.}
\label{lineR}
\end{figure}

We investigated any radial gradient of emission line ratios of M81 SNRs.
In {\bf Figure \ref{lineR}} we displayed four line ratios (\Othree $\lambda$5007/\Hbb, \Othree $\lambda$5007/\Otwo, \Ntwo/\Hab, and \Stwo/\Ha ratios) versus galactocentric distance  of M81 SNRs. 
We plotted also the data for M31 and M33 \citep{gal99, gor98}. 
Several features are noted in this figure.
First, M81 SNRs  show little radial gradients in \Othree$\lambda$5007/\Hb and  \Othree$\lambda$5007/\Otwo~ ratios. 
Second, M81 SNRs show clearly a radial gradient in \Ntwo/\Ha ratios. 
Linear fitting for 26 SNRs yields 
 Log \Ntwo/\Ha $= (-0.013\pm0.007) R + (0.085\pm0.048)$ with $rms=0.09$
 ($(-0.018\pm0.008) R +( 0.113\pm0.055)$ with $rms=0.09$ if two outliers are excluded).
Even if we select only the \Othree-strong SNRs, we obtain similar results:
 Log \Ntwo/\Ha $= (-0.015\pm0.006) R + (0.132\pm0.040)$ with $rms=0.07$.
 Similarly we derive a linear fit for 72 SNRs in M33:
 Log \Ntwo/\Ha $= (-0.050\pm0.006) R -(0.238\pm0.026)$ with $rms=0.124$.
 \citet{gal99} presented a value of the radial gradient for M31 SNRs, $-0.04\pm0.01$
dex kpc$^{-1}$. 
Thus the slope of M81 SNRs is two to three times flatter than those of 
 M31 SNRs  and  M33 SNRs. 
Third, the \Stwo/\Ha ratios of M81 SNRs also show a radial gradient. 
Linear fitting yields
 Log \Stwo/\Ha $= (-0.016\pm0.008) R + (0.104\pm0.057)$ with $rms=0.09$, 
 showing a similar slope to that of the \Ntwo/\Ha ratios.
 Even if only the \Othree-strong SNRs are used, we obtain similar results:
 Log \Stwo/\Ha $= (-0.017\pm0.008) R + (0.111\pm0.051)$ with $rms=0.08$.
 This is consistent with the tight correlation between these two line ratios, as described before. 

\begin{figure}
\centering
\includegraphics[scale=0.5]{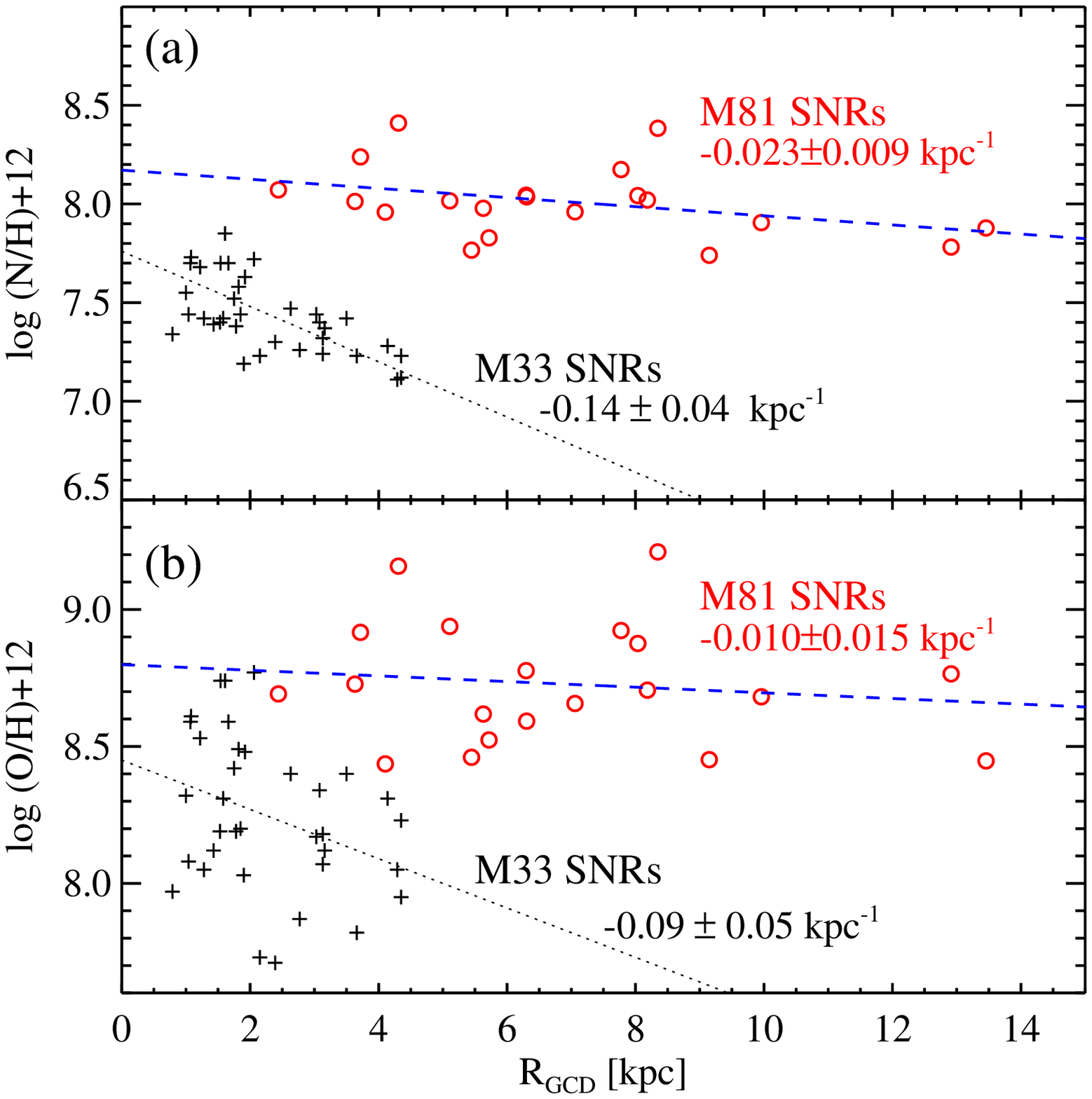}
\caption{Nitrogen (a) and oxygen (b) abundance vs. galactocentric distance of M81 SNRs (circles) in this study, 
in comparison with M33 SNRs (plusses, \citealp{smi93}). 
Note that the abundances of M81 SNRs were derived from  the \Ntwo/Ha vs. \Othree/\Ha grid in this study, 
while those of M33 SNRs were from the \Ntwo/Ha vs. \Stwo$\lambda$6731/\Ha grid by \citet{smi93}.}
\label{logONR}
\end{figure}

In {\bf Figure \ref{logONR}} we plotted Log(N/H) and Log(O/H) versus galactocentric distance of M81 SNRs 
 derived from the comparison with the \Ntwo/\Ha versus \Othree $\lambda$5007/\Hb grid of the shock ionization models in the previous section.
We also plotted the data for M33 SNRs derived from the \Ntwo/\Ha versus \Stwo $\lambda$6731/\Ha grid of the same models in \citet{smi93}.  
\citet{smi93} covered only the red wavelength in their spectra 
 so that they had no data for the \Othree~ lines and 
 could not use the \Ntwo/\Ha versus \Othree/\Ha grid for their analysis. 
Fortunately M33 SNRs have low abundance so that they could estimate the abundance of M33 SNRs using the \Ntwo/\Ha versus \Stwo $\lambda$6731/\Ha grid.
M81 SNRs show clearly a radial gradient in Log (N/H), and 
 we derived from linear fitting for 21 SNRs at $2<R<14$ kpc,
Log (N/H) $+ 12 = (-0.023\pm0.009) R + (8.154\pm0.069)$ with $rms=0.184$
 ($= (-0.020\pm0.007) R + (8.101\pm0.049)$ with $rms=0.125$, if two outliers are excluded).
This slope is much flatter than that of M33 SNRs
($0.5<R<4.5$ kpc), $-0.14\pm 0.04$ dex kpc$^{-1}$ \citep{smi93}.
On the other hand, M81 SNRs show little radial gradient in Log (O/H). 
\citet{smi93} presented a weak oxygen gradient for M33 SNRs, $-0.09\pm 0.05$ dex kpc$^{-1}$,  but with a large scatter.
These results for the abundances are consistent with the results for the line ratios for M81 SNRs.  

It is known that the pre-SN WR stars can enrich their surroundings with CN processed gas from mass-loss, as seen in the study of WR ring nebulae such as NGC 6888 \citep{mes14}.
This would raise the N abundance, while keeping S and O abundances unchanged. However, M81 SNRs show radial gradients in both \Ntwo/\Ha and \Stwo/Ha ratios as in Figure 19.
Therefore the contribution of this effect is considered to be not significant for the nitrogen gradient of M81 SNRs.

\subsection{Results for HII Regions}

We derived the physical parameters of one \Htwo region in M81 using the strong-line analysis method, following the description given in the manual for the Nebular package in IRAF \citep{sha95}.
We also applied this procedure to two M82 objects, in case that they are \Htwo regions, rather than SNRs.
We used four empirical and theoretical calibrations for deriving oxygen abundances from emission line fluxes of these objects, 
 as done for M81 \Htwo regions in \citet{pat12} and M31 HII regions in \citet{san12}:
 empirical calibrations in \citet{pil05} and \citet{bre07},
 and  
 theoretical calibrations in  \citet{kew02} and \citet{kob04}.
 Table \ref{h2abun} lists a summary of the oxygen abundance values derived in this study. 

\begin{figure}
\centering
\includegraphics[scale=0.4]{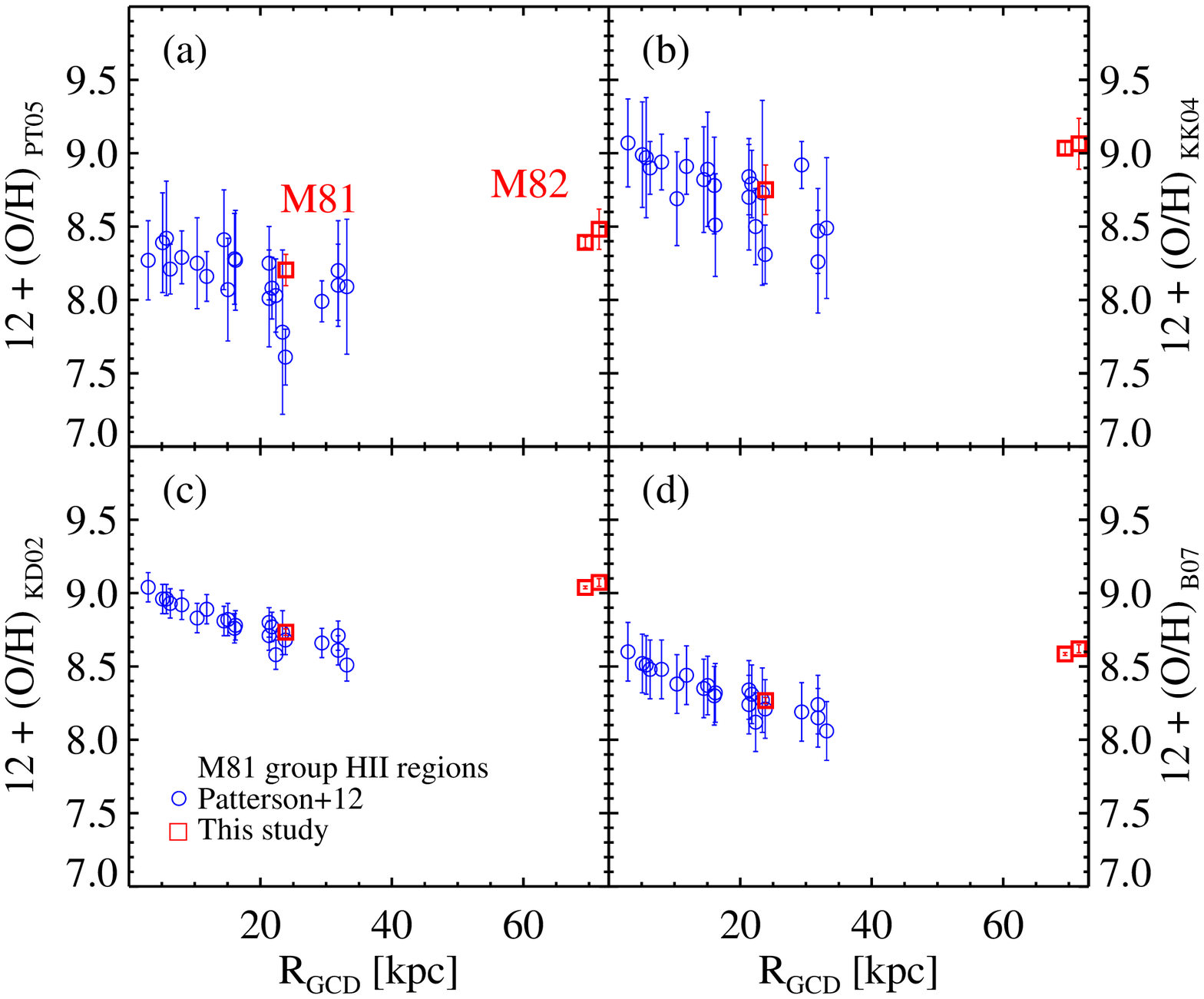} 
\caption{12+log(O/H) vs. galactocentric distance for the \Htwo regions in M81 and two M82 objects 
(red filled circles) in this study in comparison with the M81 \Htwo regions from \citet{pat12} (blue open circles). 
The outermost two objects belong to M82. 
Oxygen abundances were derived using four different calibrations: 
(a) \citet{pil05}, (b) \citet{kob04}, (c) \citet{kew02}, and (d) \citet{bre07}. }
\label{h2Ograd}
\end{figure}

{\bf Figure \ref{h2Ograd}} shows the oxygen abundances of one M81 \Htwo region we derived  versus the galactocentric distance,
 in comparison with those of other M81 \Htwo regions given by \citet{pat12}.
The value of our \Htwo region is consistent with the results of \citet{pat12} for the same galactocentric distance.
On the other hand, two objects in M82 show the oxygen abundances higher than the mean value of the M81 \Htwo regions.
\citet{smi06} derived from $HST$/STIS spectroscopy a value of the oxygen abundance for one compact \Htwo region hosting a star cluster, M82 A-1: Log(O/H)+12=9.23, 8.76, and 8.77, based on \citet{kew02}, \citet{pet04}, and adoption of a value of $T_e=10^4$ K, respectively. 
These results are close to or higher than the solar value,  Log(O/H)=8.7 to 8.9.
Thus the oxygen abundances of the two M82 objects derived in this study are similar to the value for M82 A-1.

\section{Discussion}

\subsection{Comparison of  Abundance Gradients of SNRs, \Htwo Regions, and PNe in M81}

We compared  the radial gradients of nitrogen and oxygen of SNRs derived in this study with those based on \Htwo regions and
PNe in M81 in the literature \citep{pat12,sta14}, as listed in {\bf Table \ref{gradientM81comp}}. 
\citet{pat12} analyzed a sample of \Htwo regions at the large range of $5.7 <R< 32$ kpc in M81, using the method of strong-line oxygen abundance analysis.
They derived a value for the radial oxygen gradient, $-0.013$ to $-0.020$ dex kpc$^{-1}$, 
 much flatter than the previous results based on the samples of \Htwo regions in the inner region of M81 \citep{gar87,sta10}.
 
On the other hand, \citet{sta14} derived abundances of \Htwo regions and PNe in the inner region ($ R< 13$ kpc) of M81, using the method of weak-line abundance analysis.
They suggested that the radial oxygen gradient of \Htwo regions in the inner region ($R<10$ kpc) of M81 
 is much steeper than the \citet{pat12}'s value.
Then they pointed out that there may be a break at $R\approx 15$ kpc, 
 noting that two \Htwo regions at 16 kpc and 32 kpc given by \citet{pat12} 
 showed a constant value at lower oxygen abundance, 
 log (O/H)+12 $\approx  8.1$ (see their Fig. 8).
They found also that the radial oxygen gradient of the \Htwo regions
  at $4<R<10$ kpc ($-0.088\pm 0.013$ dex kpc$^{-1}$) shows an about twice steeper slope than the radial metallicity gradient of PNe
 at $3<R<13$ kpc  ($-0.044\pm0.007$ dex kpc$^{-1}$),
 suggesting that this is due to evolution effect of metallicity gradient (see their Figure 5).
\citet{sta14} found that the radial nitrogen gradient of the \Htwo regions
 ($-0.067\pm 0.013$ dex kpc$^{-1}$) shows an about 1.5 times steeper slope than the radial metallicity gradient of PNe
 ($-0.049\pm0.007$ dex kpc$^{-1}$) (see their Figure 7).
They also pointed out that these gradients in M81 are steeper than those in other spiral galaxies (MWG, M33, and NGC 300),
 and that this may be due to its location in the galaxy group (see their Figure 9).
  The discrepancy between the results of \citet{pat12} and \citet{sta14} may be due to the differences in the method used  and the radial coverage of the samples in the two studies,  and further studies of \Htwo regions in M81 are needed to clarify this problem.

The radial gradient of 
 nitrogen abundance of the SNRs at $2<R<14$ kpc in M81 derived in this study, 
 $-0.023\pm0.009$ dex kpc$^{-1}$,
 is a few times flatter than the values for the nitrogen abundance gradient of the \Htwo regions 
 and PNe given by \citet{sta14}.
\citet{pat12} did not present any data for the nitrogen radial gradient of the \Htwo regions. 
However, the flatter oxygen radial gradient given by \citet{pat12} indicates a similarly flatter nitrogen radial  gradient, much flatter than the values given by \citet{sta14}.
Then little radial gradient of oxygen abundance
of M81 SNRs ($-0.010\pm0.015$) found in this study
may be closer to the flatter gradient given by \citet{pat12}
rather than to the steeper gradient given by \citet{sta14}.

Why SNRs show flatter radial gradients in nitrogen and oxygen than \Htwo regions and PNe in M81 is not clear, requiring further studies.
SNRs located in the spiral arms as those in this study mostly must have come from core-collapse supernovae in star-forming regions. Then they are as young as \Htwo regions. From this we expect that SNRs and \Htwo regions follow similar radial abundance gradients. 
However, optical emission lines from SNRs depend both on abundance and shock conditions, while those from \Htwo regions depend on abundance only. 
Oxygen lines in SNRs are more affected by shock conditions than nitrogen lines. Then it is expected that radial abundance gradients derived from optical spectra of SNRs and \Htwo regions will be similar for nitrogen, but different for oxygen. However, the radial gradient for nitrogen of SNRs in this study is significantly flatter than that of \Htwo regions given by \citet{sta14}. 
The cause for this discrepancy may be due to
1) the uncertainty in deriving abundance from the comparison of emission lines and shock ionization models, or
2) any evolution effect related with nitrogen.

\subsection{Comparison of  Abundance Gradients of SNRs in Nearby Galaxies}
 
In Table \ref{gradientall} we list previous estimates of the  abundance gradients based on SNRs in M31 \citep{bla85}, M33 \citep{bla85,smi93}, the Milky Way Galaxy (MWG) \citep{bla85}, as well as in M81 in this study.
M31 SNRs and M81 SNRs are similar in that they show a radial gradient in nitrogen abundance, but little in oxygen abundance, although the nitrogen gradient of M31 SNRs (--0.04) is twice steeper than that of M81.
On the other hand, M33 SNRs show radial gradients in both nitrogen and oxygen, which are steeper than those for M81 and M31. 
The MWG SNRs show a weak radial gradient of nitrogen,
similar to M81 SNRs.
There is no information available for the oxygen radial gradient of the SNRs in the MWG.

\citet{bla85} noted that the SNRs in M31 and M33 show little radial gradients and a large scatter in oxygen abundances, while they show clearly radial gradients similar to \Htwo regions but a much larger mean values in nitrogen abundance. 
They pointed out that the causes for the oxygen abundance are two: the possible confusion with low shock velocity SNRs, and the contamination due to nearby photoionized regions.
In the case of M81 SNRs in this study, we could separate the SNRs with low and high shock velocities according to the \Othree$\lambda$5007/\Hb ratios, \Othree-weak and \Othree-strong groups. 
Even if we select only the \Othree-strong SNRs, they show little radial oxygen gradients. 
As described in the previous section, the value of the \Othree $\lambda$5007/\Hb ratio changes rapidly from --0.5 at $v_s \approx 80$ \kms~ to +0.5 at  $v_s \approx 100$ \kms, and increases slowly to $\approx 0.65$ at $v_s \approx 170$ \kms, according to the model of \citet{dop84}. 
On the other hand, the values of the  \Othree$\lambda$5007/\Hb ratio for the \Othree-strong SNRs in M81 derived in this study are 0.0 to 0.7. Therefore  little radial gradients and a large scatter in oxygen abundances in M81 SNRs are mainly due to the varying shock velocity even among the \Othree-strong SNRs.
The spectra of M81 SNRs were obtained with fibers with $1\farcs5$ diameter so that it is possible that they might have been contaminated by nearby photoionized regions. However, most of M81 SNRs in this study are larger than the fiber sizes so that the effect of contamination due to nearby photoionized regions is not considered to be significant.

\subsection{Optical SNRs and X-ray Sources in M81}

\citet{pan07} found 97 X-ray sources in M81 through the Chandra X-ray Observatory (CXO) observations. 
They searched for X-ray counterparts of the optical and radio SNR candidates in M81 \citep{mat97} using this catalog, but finding none.
Later \citet{sel11} provided a catalog of 265 X-ray sources in M81 detected  in a large number of CXO fields, which is also included in \citet{liu11}. 
We crosschecked our sample of M81 SNRs with the catalog of X-ray sources in M81 given by \citet{sel11}, 
 finding five SNRs matched with X-ray sources:
 MF10=Sell259, MF17=Sell193, MF19=Sell195=Liu1578, MF25=Sell172, and L5=Sell50=Liu1653.
The images of these SNRs in Figure \ref{snrmap} show that they are compact and bright, indicating that they are relatively young.  They are all located in the inner spiral arm region, as shown in Figure 1.
They all have \Othree/\Hb $> 1$, belonging to the \Othree-strong group, which show that they have relatively high shock velocity. 

\begin{figure}
\centering
\includegraphics[scale=0.45]{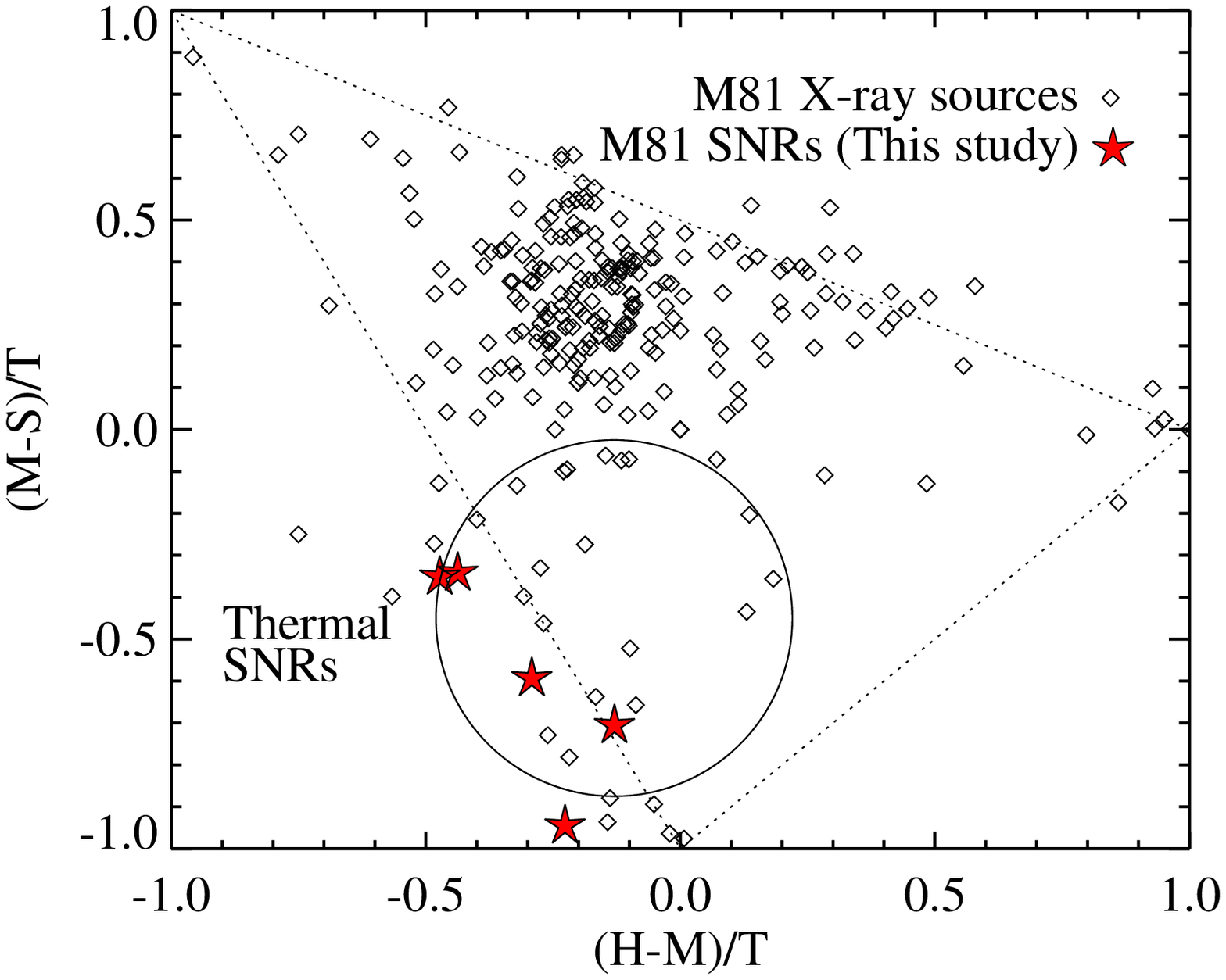} 
\caption{X-ray hardness ratio diagram (X-ray classification diagram, \citealp{pre03}) of the optical SNRs (filled starlets) matched with the catalog of M81 X-ray sources (open diamonds,  \citealp{sel11}). $S$, $M$, and $H$ represent, respectively, the counts in soft (0.5--1 keV), medium (1--2 keV), and hard (2--8 keV) bands, and $T$ for the total counts (0.5--8 keV).  Dotted lines denote the boundary for net counts.
Note that the positions of the optical SNRs are consistent with the region for thermal SNRs (large circle).
The X-ray sources with harder colors are mostly HMXBs and LMXBs.}
\label{xray}
\end{figure}

In {\bf  Figure \ref{xray}} we plot these SNRs in the X-ray hardness ratio diagram of M81 X-ray sources given by \citet{sel11}. $S$, $M$, and $H$ represent, respectively, the counts in soft (0.5--1 keV), medium (1--2 keV), and hard (2--8 keV) bands, and $T$ for the total counts (0.5--8 keV).  
 $(M-S)/T$ and $(H-M)/T$  denote  soft X-ray color and hard X-ray color, respectively.
This diagram is useful for classification of the X-ray sources in nearby galaxies to distinguish Low Mass X-ray Binaries (LMXBs), High Mass X-ray Binaries (HMXBs), SNRs, absorbed AGN sources and stars \citep{pre03}. 
X-ray sources in the high density region with hard colors are mostly HMXBs and LMXBs that are composed of either pulsar binaries or blackhole binaries and low magnetic field neutron star binaries \citep{pre03}. 
Thermal X-ray radiation is emitted from the hot plasma inside the SNRs, while non-thermal X-ray radiation comes from the shocked region in the shell regions \citep{vin12}.  The contribution of thermal line emission is strong in the medium band so that the thermal SNRs are expected to be located in the soft color region as shown in this figure.
Some of the X-ray sources located in the SNR region show variability. They are not SNRs, because SNRs do not show any variability \citep{liu11}. 

All five SNRs in M81 have soft colors and they are located in the thermal SNR region in the figure. This shows that these SNRs are all thermal SNRs.
Thus these SNRs are genuine SNRs in both optical and X-ray properties.
One of them has $(M-S)/T \approx -1.0$, which is similar to the colors of the supersoft sources. The possible origins of the supersoft sources are either SNRs or accretion-powered sources that are variable and may be  associated with young stellar population \citep{pre03}.
The presence of one optical SNR with supersoft colors in M81 provide support for the SNR origin of the supersoft sources. 

\section{Summary and Conclusions}

We obtained optical spectra of 26 SNRs and one \Htwo region in M81 and two SNRs in M82,
covering a wide range of wavelength. Analyzing the emission line fluxes of these objects in comparison with shock ionization models for SNRs, we investigated three aspects of the relations: between line ratios, between line ratios and sizes, and between line ratios as well as  abundance and galactocentric distances. We also derived oxygen abundance of one \Htwo region in M81 and two objects in M82 through strong-line analysis. Main findings are as follows.

\begin{enumerate}

\item Twenty six out of the M81 SNR candidates turn out to be genuine SNRs,
showing \Stwo/\Ha ratios larger 0.5. Two objects in M82 are shock condensations in the general outflow or SNRs.
M81 SNRs are mostly located along the spiral arms.
Their kinematics follows well the rotation of the M81 disk,
showing that all of them belong to the disk population.

\item The distribution of \Ntwo/\Ha ratios of M81 SNRs is bimodal. 
The high component is mainly due to the presence of the SNRs at the inner region ($R<5$ kpc). 

\item M81 SNRs are divided into two groups in the shock-ionization regions of the spectral line ratio diagrams: an \Othree-weak group (\Othree/\Hb$<1$) and an \Othree-strong group (\Othree/\Hb$>1$). 
The \Othree-weak SNRs may have fast shock velocity and a low fraction of shock, 
and they are relatively large. 
This implies that they may be dynamically old.

\item \Ntwo/\Ha and \Stwo/\Ha ratios of M81 SNRs show a strong correlation. These ratios can be a good calibrator of N/S abundance ratio, because these lines are coming from the same region of the shock.
\Oone/\Ha and \Stwo/\Ha ratios also show a similar correlation, but with a larger scatter.

\item \Stwo $\lambda$6717/\Stwo $\lambda$6731 ratios of the SNRs are smaller than 1.45, consistent with the theoretical density lower limit.  Most of them are larger than 1.3, indicating that the SNRs are in the low density region. 

\item \Ntwo/\Hab, \Stwo/\Hab, and \Stwo $\lambda$6717/\Stwo $\lambda$6731 ratios of the SNRs show little dependence on the SNR size.

\item \Ntwo/\Ha and \Stwo/\Ha ratios of the SNRs show a clear radial gradient,  
dLog ({\Ntwo/\Hab})/dLog R $ = -0.018\pm0.008$ dex kpc$^{-1}$  and 
dLog ({\Stwo/\Hab})/dLog R $=-0.016\pm0.008$ dex kpc$^{-1}$  
for $2<R<14$ kpc. 

\item We estimated the nitrogen and oxygen abundance of the SNRs from the comparison of their line ratios with shock-ionization models. Note that the results of determination of abundances of the SNRs based on the old simple models should be considered only as an approximate guide.
We found a value for the nitrogen radial gradient, 
dLog({N/H})/dLogR $ = -0.023\pm0.009$ dex kpc$^{-1}$, and 
little gradient for oxygen. 
Little gradient of oxygen of SNRs is mainly due to the fact that oxygen lines are significantly affected by the shock velocity and M81 SNRs have a large range of shock velocity. 

\item The nitrogen abundance of the SNRs shows a few times flatter gradient than those of the \Htwo regions and PNe in M81. The difference in the radial gradients between SNRs, PNe, and \Htwo regions in M81 remain to be explained.

\item We found X-ray SNRs in M81 for the first time. 
Five of the SNRs in M81 that were confirmed from optical spectra are matched with X-ray sources detected in the CXO observations. They look mostly  compact and bright in the \Stwo~ images. They are located in the inner spiral arm region at $R<8.3$ kpc. Their X-ray hardness colors are consistent with thermal SNRs.
The X-ray hardness colors for one of them are similar to those of the supersoft X-ray sources, supporting the SNR hypothesis for the origin of the supersoft sources.

\end{enumerate}

Through this study based on high quality spectra we found several interesting features of SNRs and SNR candidates in M81 and M82, providing strong clues to
understand the statistical properties of SNRs. 
Emission line ratios of shock models for SNRs depend not only on abundances of the interstellar medium, but also on the complexity of partially radiative shocks, degree of mixing with photoionization, shock velocity, the strength of magnetic field, and grain destruction. 
However, these factors were not included in the old models used in this study, and no modern shock model grids considering all these factors are not yet available. 
Therefore the results of determination of abundances of the SNRs based on the old simple models should be considered only as an approximate guide.
These results can be improved, when the new modern shock models are available in the future.

\acknowledgements
The authors are grateful to the referee, Mike Dopita for his comments that help us improve significantly the original draft. The authors thank to Dr.Jinyoung Serena Kim (at the University of Arizona) for helping the reduction of MMT data.
This work was supported by the National Research Foundation of Korea (NRF) grant 
by the Korea Government (MSIP)
(No. 2012R1A4A1028713).
This work was supported by K-GMT Science Program (PID: 14MMT004) funded through Korea GMT Project operated by Korea Astronomy and Space Science Institute (KASI).


\clearpage


\clearpage
\begin{turnpage}
\begin{deluxetable*}{lcllllllllll}
\tablecolumns{12}
\tabletypesize{\scriptsize}
\tablewidth{0pt}
\setlength{\tabcolsep}{0.05in}
\tablecaption{Emission-line Strengths of Emission-line Objects in M81 and M82}
\tablehead{ \multicolumn{1}{c}{ID} & 
\colhead{c(H$\beta$)} & 
\colhead{H$\beta$\tablenotemark{a} } & 
\colhead{[O II]3727,9\tablenotemark{b}} & \colhead{[O III]4959\tablenotemark{b}} & \colhead{[O III]$\lambda$5007\tablenotemark{b}} & 
\colhead{[O I]6300\tablenotemark{b}} & \colhead{[N II]6548\tablenotemark{b}} & \colhead{H$\alpha 6563$\tablenotemark{b}} & 
\colhead{[N II]6583\tablenotemark{b}} & \colhead{[S II]6717\tablenotemark{b}} & \colhead{[S II]6731\tablenotemark{b}} }
\startdata
\multicolumn{1}{l}{      L01} & $0.42 \pm0.04$ & $  5.9 \pm   0.6$ & $ 380.2 \pm  20.3$ & $ 17.9 \pm   2.6$ & $ 54.3 \pm   3.0$ & $ 17.8 \pm   3.9$ & $ 58.7 \pm   1.9$ & $321.6 \pm  10.2$ & $175.1 \pm   5.8$ & $138.7 \pm   5.1$ & $100.2 \pm   3.9$  \\
\multicolumn{1}{l}{      L02} & $0.41 \pm0.06$ & $  3.8 \pm   0.5$ & $ 390.6 \pm  26.1$ & $  9.4 \pm   3.7$ & $ 27.4 \pm   3.4$ & $ 20.9 \pm   7.1$ & $ 61.6 \pm   2.6$ & $321.1 \pm  13.1$ & $184.1 \pm   7.9$ & $121.2 \pm   5.9$ & $ 90.6 \pm   4.8$  \\
\multicolumn{1}{l}{      L03} & $0.59 \pm0.07$ & $  0.3 \pm   0.1$ & $ 368.0 \pm  31.2$ & $ 50.6 \pm   4.7$ & $131.6 \pm   5.0$ & $ 14.9 \pm   5.0$ & $ 68.7 \pm   3.3$ & $330.5 \pm  15.7$ & $204.8 \pm  10.0$ & $124.2 \pm   6.8$ & $ 93.7 \pm   5.5$  \\
\multicolumn{1}{l}{      L04} & $0.46 \pm0.04$ & $  3.6 \pm   0.3$ & $ 422.6 \pm  17.0$ & $ 25.7 \pm   2.6$ & $ 58.6 \pm   2.5$ & $ 34.8 \pm   4.4$ & $ 63.5 \pm   1.7$ & $323.4 \pm   8.4$ & $189.5 \pm   5.1$ & $142.6 \pm   4.3$ & $ 98.8 \pm   3.2$  \\
\multicolumn{1}{l}{      L05\tablenotemark{*}} & $0.47 \pm0.03$ & $  5.8 \pm   0.4$ & $ 674.6 \pm  19.1$ & $106.8 \pm   2.3$ & $321.0 \pm   3.1$ & $ 57.4 \pm   2.5$ & $ 98.4 \pm   2.1$ & $324.2 \pm   6.9$ & $293.7 \pm   6.2$ & $175.0 \pm   4.2$ & $151.1 \pm   3.7$  \\
\multicolumn{1}{l}{      L06} & $0.21 \pm0.05$ & $ 12.9 \pm   1.7$ & $ 393.1 \pm  25.0$ & $ 91.0 \pm   4.4$ & $262.4 \pm   5.4$ & $ 60.0 \pm   8.7$ & $ 54.7 \pm   2.4$ & $310.4 \pm  12.6$ & $163.7 \pm   7.1$ & $102.5 \pm   5.3$ & $ 73.0 \pm   4.2$  \\
\multicolumn{1}{l}{     MF01} & $0.45 \pm0.07$ & $  1.5 \pm   0.2$ & $ 492.9 \pm  35.6$ & $110.7 \pm   5.4$ & $336.4 \pm   6.7$ & $ 34.6 \pm  10.0$ & $ 72.0 \pm   3.6$ & $323.0 \pm  15.7$ & $215.0 \pm  10.8$ & $153.7 \pm   8.5$ & $108.0 \pm   6.4$  \\
\multicolumn{1}{l}{     MF04} & $0.59 \pm0.11$ & $  0.5 \pm   0.1$ & $ 511.9 \pm  58.1$ & $ 55.9 \pm   8.3$ & $149.0 \pm   8.6$ & $ 38.6 \pm  16.4$ & $ 74.7 \pm   5.7$ & $330.6 \pm  25.1$ & $222.7 \pm  17.4$ & $172.4 \pm  14.6$ & $137.4 \pm  12.1$  \\
\multicolumn{1}{l}{     MF05} & $0.38 \pm0.08$ & $  0.9 \pm   0.2$ & $ 601.3 \pm  44.9$ & $ 44.4 \pm   5.9$ & $134.1 \pm   6.3$ & $ 69.6 \pm   9.7$ & $ 62.5 \pm   3.6$ & $319.4 \pm  17.6$ & $186.6 \pm  10.8$ & $141.8 \pm   9.1$ & $ 98.7 \pm   7.0$  \\
\multicolumn{1}{l}{     MF07} & $0.62 \pm0.11$ & $  0.5 \pm   0.1$ & $1103.0 \pm 107.4$ & $ 64.8 \pm   9.1$ & $197.1 \pm   9.4$ & $149.1 \pm  20.1$ & $ 88.4 \pm   7.2$ & $332.4 \pm  26.9$ & $263.4 \pm  21.6$ & $220.3 \pm  19.4$ & $160.2 \pm  14.6$  \\
\multicolumn{1}{l}{     MF08} & $0.74 \pm0.17$ & $  0.0 \pm   0.0$ & $1549.1 \pm 221.0$ & $132.8 \pm  17.0$ & $282.9 \pm  15.7$ & $ 40.7 \pm  11.1$ & $121.1 \pm  14.7$ & $338.7 \pm  42.0$ & $359.4 \pm  44.5$ & $252.0 \pm  33.6$ & $180.9 \pm  25.1$  \\
\multicolumn{1}{l}{     MF10\tablenotemark{*}} & $0.67 \pm0.07$ & $  0.9 \pm   0.1$ & $ 877.4 \pm  52.4$ & $ 75.0 \pm   5.0$ & $233.8 \pm   5.9$ & $107.8 \pm   9.3$ & $ 95.3 \pm   4.5$ & $335.2 \pm  15.9$ & $283.9 \pm  13.7$ & $202.0 \pm  10.6$ & $150.7 \pm   8.3$  \\
\multicolumn{1}{l}{     MF11} & $1.01 \pm0.16$ & $  0.1 \pm   0.0$ & $1777.0 \pm 224.5$ & $ 82.2 \pm  13.8$ & $196.0 \pm  12.2$ & $136.1 \pm  19.4$ & $116.9 \pm  12.7$ & $354.0 \pm  38.8$ & $346.7 \pm  38.3$ & $261.4 \pm  30.8$ & $182.9 \pm  22.1$  \\
\multicolumn{1}{l}{     MF12} & $0.65 \pm0.08$ & $  0.5 \pm   0.1$ & $1214.3 \pm  85.2$ & $170.5 \pm   7.3$ & $483.8 \pm   9.2$ & $ 38.4 \pm   8.8$ & $ 95.8 \pm   5.6$ & $333.9 \pm  19.5$ & $285.4 \pm  16.9$ & $198.1 \pm  12.8$ & $139.3 \pm   9.5$  \\
\multicolumn{1}{l}{     MF16} & $0.74 \pm0.11$ & $  0.6 \pm   0.2$ & $1043.2 \pm  96.8$ & $ 84.2 \pm   9.0$ & $197.2 \pm   9.0$ & $ 47.9 \pm  13.0$ & $ 76.5 \pm   5.8$ & $338.7 \pm  25.2$ & $227.8 \pm  17.4$ & $188.6 \pm  15.5$ & $131.0 \pm  11.3$  \\
\multicolumn{1}{l}{     MF17\tablenotemark{*}} & $0.50 \pm0.04$ & $  2.8 \pm   0.3$ & $ 709.6 \pm  28.5$ & $ 58.1 \pm   3.1$ & $167.3 \pm   3.6$ & $ 92.3 \pm   5.3$ & $104.7 \pm   3.2$ & $325.4 \pm  10.0$ & $312.3 \pm   9.6$ & $234.3 \pm   7.8$ & $172.4 \pm   6.0$  \\
\multicolumn{1}{l}{     MF19\tablenotemark{*}} & $0.54 \pm0.03$ & $  3.7 \pm   0.3$ & $ 655.1 \pm  21.6$ & $ 75.4 \pm   2.5$ & $215.5 \pm   3.1$ & $ 70.4 \pm   4.2$ & $ 96.2 \pm   2.3$ & $327.8 \pm   7.9$ & $286.8 \pm   6.9$ & $173.0 \pm   4.7$ & $148.9 \pm   4.1$  \\
\multicolumn{1}{l}{     MF21} & $0.34 \pm0.05$ & $  7.6 \pm   0.9$ & $ 583.4 \pm  29.1$ & $ 37.3 \pm   3.2$ & $135.8 \pm   4.1$ & $ 94.3 \pm   5.8$ & $ 82.6 \pm   3.0$ & $317.3 \pm  11.4$ & $246.8 \pm   9.0$ & $179.2 \pm   7.2$ & $137.3 \pm   5.8$  \\
\multicolumn{1}{l}{     MF22} & $0.82 \pm0.13$ & $  0.2 \pm   0.0$ & $ 664.7 \pm  87.8$ & $  7.4 \pm   6.5$ & $ 61.8 \pm   9.0$ & $113.9 \pm  16.3$ & $ 78.3 \pm   7.0$ & $343.5 \pm  30.6$ & $232.8 \pm  21.3$ & $188.9 \pm  18.5$ & $137.6 \pm  14.0$  \\
\multicolumn{1}{l}{     MF25\tablenotemark{*}} & $0.69 \pm0.06$ & $  0.7 \pm   0.1$ & $ 904.7 \pm  48.8$ & $ 91.2 \pm   4.7$ & $260.9 \pm   5.4$ & $ 75.1 \pm   7.2$ & $ 99.8 \pm   4.2$ & $335.9 \pm  14.2$ & $297.2 \pm  12.6$ & $194.0 \pm   9.0$ & $155.4 \pm   7.5$  \\
\multicolumn{1}{l}{     MF26} & $0.88 \pm0.16$ & $  0.0 \pm   0.0$ & $ 917.8 \pm 133.2$ & $ 58.1 \pm  15.1$ & $109.4 \pm  13.3$ & $ 68.4 \pm  20.1$ & $102.6 \pm  11.6$ & $346.9 \pm  39.2$ & $304.7 \pm  34.8$ & $233.4 \pm  28.5$ & $169.5 \pm  21.5$  \\
\multicolumn{1}{l}{     MF27} & $0.58 \pm0.07$ & $  0.4 \pm   0.1$ & $ 712.4 \pm  49.4$ & $101.7 \pm   6.0$ & $303.5 \pm   7.2$ & $ 35.4 \pm   6.5$ & $ 83.5 \pm   4.4$ & $329.8 \pm  17.3$ & $248.8 \pm  13.3$ & $164.6 \pm   9.8$ & $124.4 \pm   7.8$  \\
\multicolumn{1}{l}{     MF29} & $0.57 \pm0.06$ & $  1.5 \pm   0.2$ & $ 363.8 \pm  28.6$ & $ 19.0 \pm   5.6$ & $ 27.6 \pm   4.3$ & $ 34.6 \pm   7.2$ & $ 53.8 \pm   2.5$ & $329.5 \pm  14.6$ & $160.5 \pm   7.6$ & $132.8 \pm   6.9$ & $ 91.0 \pm   5.1$  \\
\multicolumn{1}{l}{     MF32} & $0.53 \pm0.10$ & $  0.7 \pm   0.2$ & $1022.7 \pm  92.6$ & $ 56.5 \pm   8.4$ & $183.0 \pm   9.2$ & $150.1 \pm  15.7$ & $ 93.2 \pm   7.0$ & $327.3 \pm  24.7$ & $278.0 \pm  21.2$ & $208.7 \pm  17.3$ & $143.7 \pm  12.6$  \\
\multicolumn{1}{l}{     MF33} & $0.75 \pm0.13$ & $  0.6 \pm   0.2$ & $1866.7 \pm 189.4$ & $190.2 \pm  12.0$ & $419.5 \pm  12.3$ & $ 73.9 \pm  14.1$ & $112.4 \pm   9.9$ & $339.4 \pm  30.3$ & $334.4 \pm  30.0$ & $237.2 \pm  22.9$ & $182.0 \pm  18.0$  \\
\multicolumn{1}{l}{     MF35} & $0.71 \pm0.05$ & $  3.4 \pm   0.4$ & $ 484.3 \pm  26.0$ & $ 16.6 \pm   3.2$ & $ 41.0 \pm   3.1$ & $108.5 \pm   6.1$ & $ 65.6 \pm   2.2$ & $337.1 \pm  11.0$ & $195.4 \pm   6.6$ & $210.0 \pm   7.5$ & $156.3 \pm   5.8$  \\
\cline{1-12}
\multicolumn{1}{l}{   M82-L1} & $0.68 \pm0.03$ & $  1.8 \pm   0.1$ & $ 208.8 \pm   9.7$ & $ 20.0 \pm   1.6$ & $ 53.6 \pm   1.6$ & $ 14.3 \pm   1.8$ & $ 47.3 \pm   1.1$ & $313.3 \pm   8.1$ & $140.7 \pm   3.2$ & $ 59.9 \pm   1.7$ & $ 46.3 \pm   1.5$  \\
\multicolumn{1}{l}{   M82-L2} & $0.57 \pm0.14$ & $  5.8 \pm   1.9$ & $ 171.8 \pm  19.4$ & $ 18.7 \pm   1.6$ & $ 55.1 \pm   1.7$ & $ 16.9 \pm   2.5$ & $ 47.1 \pm   4.6$ & $307.7 \pm  47.1$ & $140.4 \pm  13.8$ & $ 51.1 \pm   5.4$ & $ 41.9 \pm   4.5$  \\
\cline{1-12}
\multicolumn{1}{l}{M81HII-L1} & $0.46 \pm0.05$ & $ 19.4 \pm   2.2$ & $ 353.9 \pm  21.0$ & $ 55.7 \pm   3.3$ & $165.2 \pm   3.8$ & $  2.6 \pm   1.9$ & $ 19.7 \pm   0.9$ & $301.9 \pm  10.0$ & $ 58.8 \pm   2.8$ & $ 24.4 \pm   2.1$ & $ 16.4 \pm   1.9$  \\
\enddata
\tablenotetext{*}{X-ray detected sources \citep{sel11}.}
\tablenotetext{a}{H$\beta$ flux in units of 10$^{-17} {\rm erg ~cm}^{-2} {\rm ~s}^{-1} {\rm ~\AA}^{-1}$. }
\tablenotetext{b}{All other fluxes are normalized by H$\beta = 100$.}
\label{line}
\end{deluxetable*}
\end{turnpage}

\begin{turnpage}
\begin{deluxetable*}{lllcccccccccc}
\tablecolumns{13}
\tabletypesize{\scriptsize}
\tablewidth{0pt}
\setlength{\tabcolsep}{0.05in}
\tablecaption{Line Ratios and Abundances of Emission-line Objects in M81 and M82}
\tablehead{\multicolumn{1}{l}{ID} & \multicolumn{1}{l}{Size} & 
\colhead{R } & \colhead{$v$} & 
\colhead{R.A.} & \colhead{Dec.} & 
\multirow{2}{*}{$\frac{{\rm [OIII]} \lambda5007}{{\rm H}\beta}$} & 
\multirow{2}{*}{$\frac{{\rm [O I]}6300}{{\rm H}\alpha}$} & 
\multirow{2}{*}{$\frac{{\rm [N II]}6583}{{\rm H}\alpha}$} & 
\multirow{2}{*}{$\frac{{\rm [S II]}6717,31}{{\rm H}\alpha}$} & 
\multirow{2}{*}{$\frac{{\rm [S II]}6717}{{\rm [S II]}6731}$} & 
\colhead{log (N/H)\tablenotemark{a}} & \colhead{log (O/H)\tablenotemark{a}} \\
 & [pc] & [kpc] & \colhead{${\rm km ~s}^{-1}$} & (J2000) & (J2000) &
 & & & & & & }
\startdata
\multicolumn{1}{l}{   L01} &  83.9 &  5.39 & $  82.44 \pm  2.1$ & 09:54:39.52 & 69:05:36.1 & $0.54 \pm 0.03$ & $0.06 \pm 0.01$ & $0.54 \pm 0.03$ & $0.74 \pm 0.03$ & $1.38 \pm 0.07$ & --- & ---  \\
\multicolumn{1}{l}{   L02} & 143.5 &  4.73 & $-100.43 \pm  3.3$ & 09:55:21.58 & 69:01:46.9 & $0.27 \pm 0.03$ & $0.07 \pm 0.02$ & $0.57 \pm 0.03$ & $0.66 \pm 0.04$ & $1.34 \pm 0.10$ & --- & ---  \\
\multicolumn{1}{l}{   L03} &  59.1 &  5.45 & $-135.81 \pm  2.1$ & 09:55:32.08 & 69:01:03.0 & $1.32 \pm 0.05$ & $0.04 \pm 0.02$ & $0.62 \pm 0.04$ & $0.66 \pm 0.04$ & $1.33 \pm 0.11$ &--4.23 & --3.54 \\
\multicolumn{1}{l}{   L04} &  95.3 &  6.76 & $  84.84 \pm  1.8$ & 09:55:34.10 & 69:07:29.3 & $0.59 \pm 0.03$ & $0.11 \pm 0.01$ & $0.59 \pm 0.02$ & $0.75 \pm 0.03$ & $1.44 \pm 0.06$ & --- & --- \\
\multicolumn{1}{l}{   L05} &  68.8 &  7.77 & $-199.96 \pm  0.9$ & 09:55:47.84 & 68:59:28.6 & $3.21 \pm 0.03$ & $0.18 \pm 0.01$ & $0.91 \pm 0.03$ & $1.01 \pm 0.03$ & $1.16 \pm 0.04$ &--3.83 & --3.08  \\
\multicolumn{1}{l}{   L06} & 164.9 & 12.91 & $-217.05 \pm  1.5$ & 09:56:29.29 & 68:56:16.5 & $2.62 \pm 0.07$ & $0.19 \pm 0.03$ & $0.53 \pm 0.03$ & $0.57 \pm 0.03$ & $1.40 \pm 0.11$ &--4.22 & --3.23  \\
\multicolumn{1}{l}{  MF01} &  39.7 &  5.11 & $  46.17 \pm  1.5$ & 09:54:44.27 & 69:04:23.9 & $3.36 \pm 0.07$ & $0.11 \pm 0.03$ & $0.67 \pm 0.05$ & $0.81 \pm 0.05$ & $1.42 \pm 0.12$ &--3.98 & --3.06  \\
\multicolumn{1}{l}{  MF04} &  29.7 &  5.72 & $  30.58 \pm  3.3$ & 09:54:50.95 & 69:02:57.7 & $1.49 \pm 0.09$ & $0.12 \pm 0.05$ & $0.67 \pm 0.07$ & $0.94 \pm 0.09$ & $1.25 \pm 0.15$ &--4.17 & --3.48  \\
\multicolumn{1}{l}{  MF05} &  34.0 &  9.15 & $ 171.48 \pm  2.7$ & 09:54:54.38 & 69:09:19.9 & $1.34 \pm 0.06$ & $0.22 \pm 0.03$ & $0.58 \pm 0.05$ & $0.75 \pm 0.05$ & $1.44 \pm 0.14$ &--4.26 & --3.55  \\
\multicolumn{1}{l}{  MF07} &  49.0 &  7.06 & $ 159.79 \pm  2.7$ & 09:55:00.02 & 69:08:05.5 & $1.97 \pm 0.09$ & $0.45 \pm 0.07$ & $0.79 \pm 0.09$ & $1.14 \pm 0.12$ & $1.38 \pm 0.17$ &--4.04 & --3.34  \\
\multicolumn{1}{l}{  MF08} &  29.5 &  3.72 & $ 104.93 \pm  3.3$ & 09:55:04.24 & 69:05:54.6 & $2.83 \pm 0.16$ & $0.12 \pm 0.04$ & $1.06 \pm 0.19$ & $1.28 \pm 0.20$ & $1.39 \pm 0.27$ &--3.76 & --3.08  \\
\multicolumn{1}{l}{  MF10} &  59.2 &  3.63 & $ -15.29 \pm  2.4$ & 09:55:07.17 & 69:03:14.6 & $2.34 \pm 0.06$ & $0.32 \pm 0.03$ & $0.85 \pm 0.06$ & $1.05 \pm 0.06$ & $1.34 \pm 0.10$ &--3.99 & --3.27  \\
\multicolumn{1}{l}{  MF11} &  52.5 &  2.44 & $  39.57 \pm  3.6$ & 09:55:09.38 & 69:04:14.7 & $1.96 \pm 0.12$ & $0.38 \pm 0.07$ & $0.98 \pm 0.15$ & $1.25 \pm 0.17$ & $1.43 \pm 0.24$ &--3.93 & --3.31  \\
\multicolumn{1}{l}{  MF12} &  40.9 &  8.35 & $ 175.98 \pm  1.5$ & 09:55:10.32 & 69:08:46.8 & $4.84 \pm 0.09$ & $0.12 \pm 0.03$ & $0.85 \pm 0.07$ & $1.01 \pm 0.08$ & $1.42 \pm 0.13$ &--3.62 & --2.79  \\
\multicolumn{1}{l}{  MF16} &  57.8 &  9.96 & $ 134.91 \pm  2.7$ & 09:55:19.13 & 69:09:31.5 & $1.97 \pm 0.09$ & $0.14 \pm 0.04$ & $0.67 \pm 0.07$ & $0.94 \pm 0.09$ & $1.44 \pm 0.17$ &--4.09 & --3.32  \\
\multicolumn{1}{l}{  MF17} &  87.5 &  6.30 & $ 166.98 \pm  1.5$ & 09:55:19.72 & 69:07:32.6 & $1.67 \pm 0.04$ & $0.28 \pm 0.02$ & $0.96 \pm 0.04$ & $1.25 \pm 0.05$ & $1.36 \pm 0.07$ &--3.96 & --3.41  \\
\multicolumn{1}{l}{  MF19} &  55.0 &  8.18 & $ 124.71 \pm  0.9$ & 09:55:21.55 & 69:08:31.7 & $2.16 \pm 0.03$ & $0.21 \pm 0.01$ & $0.87 \pm 0.03$ & $0.98 \pm 0.03$ & $1.16 \pm 0.04$ &--3.98 & --3.29  \\
\multicolumn{1}{l}{  MF21} &  47.6 & 13.46 & $-174.18 \pm  1.8$ & 09:55:31.90 & 68:56:47.6 & $1.36 \pm 0.04$ & $0.30 \pm 0.02$ & $0.78 \pm 0.04$ & $1.00 \pm 0.05$ & $1.31 \pm 0.08$ &--4.12 & --3.55  \\
\multicolumn{1}{l}{  MF22} &  44.9 &  6.37 & $-173.58 \pm  5.7$ & 09:55:32.27 & 69:00:33.3 & $0.62 \pm 0.09$ & $0.33 \pm 0.06$ & $0.68 \pm 0.09$ & $0.95 \pm 0.11$ & $1.37 \pm 0.19$ & --- &  --- \\
\multicolumn{1}{l}{  MF25} &  46.6 &  6.30 & $  66.85 \pm  1.5$ & 09:55:42.07 & 69:07:00.1 & $2.61 \pm 0.05$ & $0.22 \pm 0.02$ & $0.88 \pm 0.05$ & $1.04 \pm 0.06$ & $1.25 \pm 0.08$ &--3.96 & --3.22  \\
\multicolumn{1}{l}{  MF26} &  56.5 &  4.10 & $  -5.70 \pm  6.9$ & 09:55:52.15 & 69:05:21.6 & $1.09 \pm 0.13$ & $0.20 \pm 0.06$ & $0.88 \pm 0.14$ & $1.16 \pm 0.17$ & $1.38 \pm 0.24$ &--4.04 & --3.56  \\
\multicolumn{1}{l}{  MF27} &  26.6 &  8.04 & $-239.23 \pm  1.8$ & 09:55:52.25 & 68:59:16.2 & $3.04 \pm 0.07$ & $0.11 \pm 0.02$ & $0.75 \pm 0.06$ & $0.88 \pm 0.06$ & $1.32 \pm 0.11$ &--3.96 & --3.12  \\
\multicolumn{1}{l}{  MF29} &  77.7 &  8.33 & $-217.95 \pm  3.9$ & 09:56:04.49 & 68:58:60.0 & $0.28 \pm 0.04$ & $0.11 \pm 0.02$ & $0.49 \pm 0.03$ & $0.68 \pm 0.04$ & $1.46 \pm 0.11$ & --- &  --- \\
\multicolumn{1}{l}{  MF32} &  44.0 &  5.63 & $-191.87 \pm  3.3$ & 09:56:15.83 & 69:00:51.4 & $1.83 \pm 0.09$ & $0.46 \pm 0.06$ & $0.85 \pm 0.09$ & $1.08 \pm 0.10$ & $1.45 \pm 0.17$ &--4.02 & --3.38  \\
\multicolumn{1}{l}{  MF33} &  55.3 &  4.31 & $-100.13 \pm  2.1$ & 09:56:16.18 & 69:02:39.6 & $4.19 \pm 0.12$ & $0.22 \pm 0.05$ & $0.99 \pm 0.12$ & $1.24 \pm 0.14$ & $1.30 \pm 0.18$ &--3.59 & --2.84  \\
\multicolumn{1}{l}{  MF35} &  54.6 &  6.58 & $ -89.64 \pm  2.7$ & 09:56:21.77 & 69:05:00.9 & $0.41 \pm 0.03$ & $0.32 \pm 0.02$ & $0.58 \pm 0.03$ & $1.09 \pm 0.05$ & $1.34 \pm 0.07$ & --- &  --- \\
\multicolumn{1}{l}{M82-L1} &  ---  & 69.43 & $ 105.53 \pm  1.5$ & 09:55:58.75 & 69:40:09.3 & $0.54 \pm 0.02$ & $0.05 \pm 0.01$ & $0.45 \pm 0.02$ & $0.34 \pm 0.01$ & $1.29 \pm 0.06$ & --- & --- \\
\multicolumn{1}{l}{M82-L2} &  ---  & 71.53 & $ 258.72 \pm  2.7$ & 09:55:47.29 & 69:41:35.2 & $0.55 \pm 0.02$ & $0.05 \pm 0.01$ & $0.46 \pm 0.08$ & $0.30 \pm 0.05$ & $1.22 \pm 0.18$ & --- & --- \\
\multicolumn{1}{l}{M81HII-L1} &  ---  & 23.87 & $-169.98 \pm  1.5$ & 09:56:14.29 & 68:50:20.7 & $1.65 \pm 0.05$ & $0.01 \pm 0.01$ & $0.19 \pm 0.01$ & $0.14 \pm 0.01$ & $1.49 \pm 0.22$ & --- & --- \\
\enddata
\tablenotetext{a}{Abundances derived from the comparison between observation and models on the \Ntwo/\Ha--\Othree/\Ha diagram. }
\label{ratio}
\end{deluxetable*}
\end{turnpage}

\begin{deluxetable*}{lccc}
\tabletypesize{\scriptsize}
\tablecolumns{5}
\tablewidth{0pc}
\setlength{\tabcolsep}{0.05in}
\tablecaption{Abundance Gradients of SNRs, \Htwo regions, and PNe in M81}
\tablehead{\colhead{Type} & \colhead{dLog(N/H/)dR} & \colhead{dLog(O/H)/dR} & \colhead{References}}
\startdata
SNRs  			& $-0.023\pm0.009$    & $-0.010\pm0.015$     & This study    \\
\Htwo regions   & $-0.067$            & $-0.088\pm0.013$     & \citet{sta14} \\
                & --                  & $-0.013 \sim -0.020$ & \citet{pat12} \\
PNe    			& $-0.049$            & $-0.044 \pm 0.007$   & \citet{sta14}
\enddata
\label{gradientM81comp}
\end{deluxetable*}

\begin{deluxetable*}{lcccc}
\tabletypesize{\scriptsize}
\tablecolumns{5}
\tablewidth{0pt}
\setlength{\tabcolsep}{0.05in}
\tablecaption{Abundance Gradients of SNRs in Nearby Galaxies}
\tablehead{\colhead{Galaxy} & \colhead{R$_{25}$ [kpc]} & \colhead{dLog(N/H)/dR} & \colhead{dLog(O/H)/dR} & \colhead{References}}
\startdata
M81  & 18.34\tablenotemark{a} & $-0.023\pm0.009$ & $-0.010\pm0.015$  & This study   \\
M31  & 23.56\tablenotemark{b} & $-0.040$                & $+0.004$               & \citet{bla85} \\
M33  &   5.77\tablenotemark{c} & $-0.089$                & $-0.035$ 			     & \citet{bla85} \\
  &    & $-0.140\pm0.040$ & $-0.090\pm0.050$  & Smith et al. (1993) \\
MWG & 11.5\tablenotemark{d}  & $-0.09$                  & --                             & \citet{bla85} 
\enddata
\tablenotetext{a}{\citet{dev91}.}
\tablenotetext{b}{\citet{dev91}. The distance to M31 adopted is $793\pm45$ kpc \citep{bat14}.}
\tablenotetext{c}{\citet{bla85}.}
\tablenotetext{d}{\citet{bla85}.} 
\label{gradientall}
\end{deluxetable*}

\begin{deluxetable*}{lccccc}
\tablecolumns{6}
\tabletypesize{\scriptsize}
\tablewidth{0pt}
\setlength{\tabcolsep}{0.05in}
\tablecaption{Oxygen abundance of \Htwo regions}
\tablehead{
 &  & \multicolumn{4}{c}{12+log(O/H)} \\
\colhead{ID} & \colhead{R$_{{\rm GCD}}$[kpc]\tablenotemark{a}} & \colhead{PT05\tablenotemark{1}} & \colhead{KK04\tablenotemark{2}} & \colhead{KD02\tablenotemark{3}} & \colhead{B07\tablenotemark{4}}  }
\startdata
   M82-L1\tablenotemark{b} & 69.43 & $8.39 \pm 0.04$ & $9.03 \pm 0.04$ & $9.04 \pm 0.01$ & $8.58 \pm 0.01$ \\
   M82-L2\tablenotemark{b} & 71.53 & $8.48 \pm 0.14$ & $9.06 \pm 0.17$ & $9.07 \pm 0.03$ & $8.62 \pm 0.03$ \\
M81HII-L1 & 23.87 & $8.20 \pm 0.11$ & $8.75 \pm 0.17$ & $8.73 \pm 0.02$ & $8.27 \pm 0.02$ 
\enddata
\tablenotetext{a}{Deprojected distance from the center of M81.}
\tablenotetext{b}{Abundances were derived in the case that M82 objects are fully photo-ionized.}
\tablenotetext{1}{Based on the calibration of \citet{pil05}.}
\tablenotetext{2}{Based on the calibration of \citet{kob04}.}
\tablenotetext{3}{Based on the calibration of \citet{kew02}.}
\tablenotetext{4}{Based on the calibration of \citet{bre07}.}
\label{h2abun}
\end{deluxetable*}
\clearpage

\end{document}